\PassOptionsToPackage{table}{xcolor}
\documentclass[sigconf]{acmart}
 
\AtBeginDocument{%
  }

\usepackage{float}
\usepackage{enumitem}
\usepackage{listings}
\usepackage{color}
\usepackage{soul}    
\usepackage{gensymb} 
\usepackage{subcaption}
\usepackage{graphicx}
\usepackage{enumitem} 
\usepackage{tabularray}
\usepackage{listings}
\usepackage{gensymb} 
\usepackage{xspace} 
\usepackage{booktabs}
\usepackage{caption}
\usepackage{tabularx}
\usepackage{array}          
\renewcommand{\arraystretch}{1.4}  

\usepackage[utf8]{inputenc}
\usepackage{multirow}
\usepackage{array}
\usepackage{ragged2e}

\lstdefinestyle{prompt}{
  basicstyle=\ttfamily\small\color{black},
  showstringspaces=false,
  breaklines=true
}

\newcommand{\app}{\textsc{UXCascade}\xspace}
\newcommand{\icon}[1]{\raisebox{-0.2em}{\includegraphics[height=2.5ex]{#1}}}
\renewcommand{\arraystretch}{1.6} 

\definecolor{None}{HTML}{E5E5E5} 
\definecolor{Baseline}{HTML}{A8D8D4} %
\definecolor{UXCascade}{HTML}{D8A8A8} 
\newcommand{\colUX}[1]{\cellcolor{UXCascade}{#1}}
\newcommand{\colNo}[1]{\cellcolor{None}{#1}}
\newcommand{\colBase}[1]{\cellcolor{Baseline}{#1}}

\newcommand{\changes}[1]{\textcolor{black}{#1}}

\newcommand{\revision}[1]{{\color{black}#1}}

\definecolor{listingbg}{RGB}{245,245,245} 

\copyrightyear{2026}
\acmYear{2026}
\setcopyright{cc}
\setcctype{by-nc-nd}
\acmConference[UIST '26]{The 39th Annual ACM Symposium on User Interface Software and Technology}{November 02--05, 2026}{Detroit, MI, USA}
\acmBooktitle{The 39th Annual ACM Symposium on User Interface Software and Technology (UIST '26), November 02--05, 2026, Detroit, MI, USA}
\acmDOI{10.1145/3830398.3830609}
\acmISBN{979-8-4007-2856-3/2026/11}

\begin{document}

\title[\texorpdfstring{\app: Scalable Usability Testing with Simulated Agents}{\app}]{\app: Scalable Usability Testing with \\ Simulated User Agents}


\author{Steffen Holter}
\orcid{0009-0008-2935-5549}
\email{steffen.holter@inf.ethz.ch}
\authornote{This work was done while the author was an intern at Adobe Research.}
\affiliation{%
  \institution{ETH Zurich}
  \city{Zurich}
  \country{Switzerland}
}

\author{Eunyee Koh}
\orcid{0000-0003-2091-5972}
\email{eunyee@adobe.com}
\affiliation{%
  \institution{Adobe Research}
  \city{San Jose}
  \state{CA}
  \country{USA}
}

\author{Mustafa Doga Dogan}
\orcid{0000-0003-3983-1955}
\email{doga@adobe.com}
\authornote{The authors contributed equally and jointly supervised this work.}
\affiliation{%
 \institution{Adobe Research}
 \city{Basel}
 \country{Switzerland}
}

\author{Gromit Yeuk-Yin Chan}
\orcid{0000-0003-1356-4406}
\email{ychan@adobe.com}
\authornotemark[2]
\affiliation{%
  \institution{Adobe Research}
  \city{San Jose}
  \state{CA}
  \country{USA}
}
\renewcommand{\shortauthors}{Holter et al.}

\begin{abstract}

Simulated user agents are increasingly deployed in usability testing to support fast, iterative UX workflows, as they generate rich data such as action logs and think-aloud reasoning, but the unstructured nature of this output often obscures actionable insights. We present \app, an interactive tool for extracting, aggregating, and presenting agent-generated usability feedback at scale. Our core contribution is a multi-level analysis workflow that (1) highlights patterns across persona traits, goals, and outcomes, (2) links agent reasoning to specific issues, and (3) supports actionable design improvements. \app operationalizes this approach by listing agent goals, traits, and issues in a structured overview. Practitioners can explore detailed reasoning traces and annotated views, propose interface edits, and assess their impact across personas. This enables a top-down, exploration-driven analysis from patterns to concrete, actionable UX interventions. A user study with eight UX professionals demonstrates that \app integrates into existing workflows, enabling iterative feedback during early-stage interface development.

\end{abstract}



\begin{teaserfigure}
  \includegraphics[width=\textwidth]{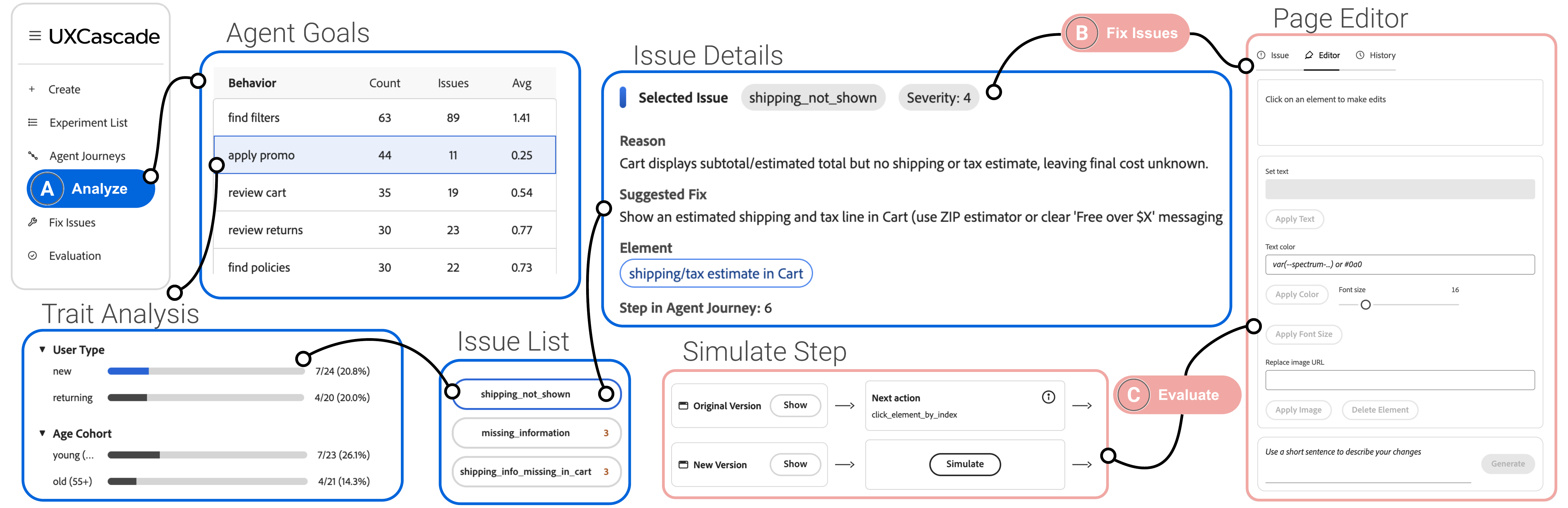}
  \caption{
  \app is an end-to-end system for simulating usability testing at scale. The workflow begins with simulated agents browsing a website to uncover issues based on user-defined goals, which are then (A) aggregated and visualized in the analysis view. Users can (B) apply interface changes to address these issues, and the system (C) automatically re-evaluates the new version by replaying agent behavior. This loop enables targeted, iterative refinement grounded in agent reasoning and interaction. \looseness=-1
  }
  \Description{
    This figure shows the main UXCascade interface, divided into three parts of the workflow.
    (A) displays the ``Agent Goals'' and ``Trait Analysis'' views, where users can examine which goals were attempted by simulated agents, how frequently issues occurred, and how these are distributed across different user traits such as age or price sensitivity.
    (B) is the editor panel, allowing users to apply interface changes by editing text, colors, and images using natural language commands.
    (C) presents the evaluation interface, enabling users to simulate the next likely user action under both the original and modified interface and generate a difference report.
  }
  \label{fig:teaser}
\end{teaserfigure}


\maketitle


\section{Introduction}

The pace of web development has never been faster, with its cycles becoming increasingly more \revision{agentic} and iterative~\cite{wang_2026_remix, wang_2026_xagen}. 
AI-assisted workflows enable user experience (UX) professionals to generate interface variations within seconds, opening opportunities to explore a far broader solution space than before~\cite{khan_2025_beyond, swearngin_2020_scout, zhou_2023_does}.
As a result, the number of interfaces needing review has outpaced the capacity of standard evaluation practices leaving many screens untested~\cite{kuang_2022_merging, folstad_2012_analysis, dierk_2025_evaluating}.
This acceleration invites us to rethink how usability testing is conducted.
Instead of treating evaluation as a slow, separate phase, we can envision methods that scale with rapid generation and provide feedback early and often.

The fundamental challenge is that anticipating usability and engagement issues is inherently difficult even for seasoned experts~\cite{nielsen_1990_heuristic, hertzum_2001_evaluator}.
Testing with human participants is therefore invaluable in theory, yet often unfeasible in practice because of time and resource constraints.
Simulation aims to bridge this gap by using AI agents as proxies for human behavior~\cite{park_2023_generative, argyle_2023_out}.
Because of its cost efficiency and reproducibility, user simulation has become a promising complement to the user-centric evaluation of websites and interfaces.
Large Language Models (LLMs) can be prompted to act as autonomous agents that resemble real users as they navigate web pages, verbalize goals, and produce step-by-step rationales~\cite{park_2023_generative, he_2024_webvoyager, ma_2023_laser}.
Recent advances have endowed  these agents with explicit personas and characteristics to inject realistic behavioral diversity into testing scenarios. 

However, simulating user populations produces abundant and verbose outputs, including reasoning traces, interaction logs, and summaries.
These unstructured artifacts contain rich insights, yet they are tedious to review and difficult to act upon~\cite{lu_2025_uxagent}.
Realizing the promise of simulation requires methods that synthesize the behavior of \textit{many} personalized agents into concise, actionable findings about usability and engagement.

In this paper, we present \app, an interactive system that \textbf{structures}, \textbf{aggregates}, and \textbf{presents} agent-generated feedback \textbf{\textit{at scale}}. 
Our solution implements a multi-stage exploratory workflow to identify persona-grounded user goals, isolate key usability issues, and iterate on potential fixes. 
More specifically, by tightly integrating agent reasoning contexts, user journey representations, and on-the-spot editing, we help practitioners move easily from raw agent logs to actionable findings.

Our approach caters to UX experts and their iterative cycles of interface design and prototyping. 
The aim is to provide an additional option for early-stage evaluation to complement established human-centered methods. 
By offering a lightweight simulation-based assessment when rapid feedback is needed, we allow user studies to be reserved for contexts where they add the most value. \looseness=-1



We evaluate our system by conducting a user study with 8 UX professionals who compare \app to human-generated feedback on a sample website seeded with usability issues. 
The results show that our solution performs comparably to the baseline condition, suggesting its potential utility as part of UX practitioners' workflows.
Overall, our main contributions are as follows:
\vspace{-0.1cm}
\begin{enumerate}
    \item\noindent \textbf{A workflow}: A five-stage workflow that groups intentions, compares personas, distills issues, and supports edit-in-the-loop iteration.
    \item\noindent \textbf{A system}: An interactive system, \app, instantiates the workflow to generate agent runs, structure and aggregate their outputs, and link reasoning to interface elements with provenance to produce actionable findings.\looseness=-1
    \item\noindent \textbf{An evaluation}: A within-subjects study with eight UX professionals using a custom website seeded with usability issues, evaluating issue discovery rates, subjective workload, and fit with existing UX workflows.
\end{enumerate}





\vspace{-0.5em}
\section{Related Work}

In this section, we review related work focusing on usability testing, simulated agents, and persona-based simulations. 

\subsection{Usability Testing}

Usability testing remains a foundational method in UX research, offering systematic insight into how users interact with digital interfaces to accomplish their goals~\cite{dumas_1993_practical, rubin_2008_handbook}. 
It supports iterative improvement by revealing interaction breakdowns, validating design decisions, and ensuring alignment with user needs. 
However, traditional usability workflows face persistent bottlenecks in experiment design and participant recruitment~\cite{alshamari_2008_task}, particularly during early-stage prototyping where timely feedback is most critical.
In practice, teams often fall back on expert inspection methods such as heuristic evaluation and cognitive walkthroughs~\cite{nielsen_1990_heuristic, lewis_1997_cognitive}, but these too depend on scarce expert time~\cite{hertzum_2001_evaluator}, motivating scalable early-stage feedback that requires neither participants nor experts.\looseness=-1

\subsection{Simulated Agents}

LLM-based agents have shown potential in simulating human-like interaction patterns within structured environments. 
Unlike deterministic scripts, these agents can reflect intentions, preferences, and domain knowledge when appropriately prompted. 
Park et al.~\cite{park_2023_generative} showed how emergent social behavior arises in a sandbox of agents with persistent memory. 
Taeb et al.~\cite{taeb_2024_axnav} built \textit{AXNav}, which translates accessibility evaluation into executable interaction sequences.\looseness=-1

Closer to usability testing, \textit{UXAgent}~\cite{lu_2025_uxagent} simulates think-aloud UX studies but focuses on protocol design rather than issue discovery.
\textit{SimUser}~\cite{xiang_2024_simuser} pairs LLMs to simulate mobile user-app interaction while \textit{USimAgent}~\cite{zhang_2024_usimagent} explores search behavior via simulated users.
Wang et al.~\cite{wang_2025_user} propose modular agent architectures with memory and profile components, enabling more realistic behavioral modeling in recommender settings.

Building on this foundation, recent work explores how LLM agents can directly automate aspects of usability evaluation.
Turbeville et al.~\cite{turbeville_2024_llmpowered} demonstrate multi-modal UX analysis using LLMs, while others have aimed to generate structured feedback on static mockups~\cite{duan_2024_generating, duan_2023_towards}.
Lu et al.~\cite{lu_2025_agentab} and Rieder et al.~\cite{rieder_2026_simab}
present techniques
for automated A/B testing using LLM agents defined via demographic and behavioral parameters.
These systems aim to lower the effort needed to validate design hypotheses, but most emphasize outcome evaluation rather than supporting iterative exploration or explaining the root causes of usability issues.
In contrast, Suh et al.~\cite{suh_2025_storyensemble} highlight the value of involving AI in dynamic, iterative design workflows through an in-depth formative study.

Still, questions of fidelity remain.
Lu et al.~\cite{lu_2025_prompting} show that LLM agents often produce unrealistic behavior without careful prompting or alignment to fine-grained data.
Gui et al.~\cite{gui_2023_challenge} caution against applying human-subject protocols uncritically to LLMs and advocate for evaluation methods tailored to generative agents.
These studies highlight the need for more reliable simulation frameworks that are both controllable and interpretable.

\subsection{Persona-Guided Simulation}
LLMs have also been shown to adopt instructed roles through prompt design (e.g., \textit{"Act as a designer"} or \textit{"You are an expert in..."})~\cite{salewski_2023_context}, a pattern exploited by several works to induce agent personas and professional perspectives.
Kosinski~\cite{kosinski_2024_theory} further argues that LLMs can infer and simulate psychological traits from textual input, suggesting their potential viability for replicating user diversity at scale.
This capacity for persona induction has already been applied across adjacent UX domains: \textit{PosterMate}~\cite{shin_2025_poster} uses persona agents for poster design, Benharrak et al.~\cite{benharrak_2024_writer} provide persona-grounded writing feedback, and \textit{Proxona}~\cite{choi_2025_proxona} generates dynamic audience personas for video content. \looseness=-1

However, capturing diverse, realistic behavior remains a central challenge in user simulation research.
Common strategies involve conditioning LLMs on demographics and goals \cite{jiang_2023_evaluating, moon_2024_virtual}, utilizing persona-labeled datasets like OPeRA \cite{wang_2025_opera}, or mining synthetic personas from behavioral logs \cite{mansour_2025_paars}.
To improve scalability, researchers have also explored reusable "carriers" – compact embeddings that encode a user's professional background and preferences \cite{ge_2024_scaling}.
Despite these advances, "naive" persona prompting often yields superficial variety rather than deep behavioral shifts \cite{aher_2023_using}.
Beyond functional issues, emotional responses are an increasingly recognized dimension of user experience~\cite{agarwal_2009_beyond} that current LLM-based systems are only beginning to address.
Furthermore, existing systems typically focus on narrow tasks, leaving the discovery of nuanced interface issues and affective responses underexplored.
Our work addresses these gaps by providing a scalable framework that grounds agent simulations in reasoning traces to support iterative UX analysis.

\section{Requirements Analysis}
\label{sec:requirements} 

In this section, we derive the design requirements for our workflow and interface from a formative study with UX professionals

\subsection{Formative Study}


\subsubsection*{Participants}
We recruited five UX professionals (2 men, 3 women; mean = 12 years of experience) from a major international software company specializing in digital design and user experience tools. 
Two participants primarily focused on UX research, while the remaining three worked as UX designers. 
In the following discussion, we refer to the experts as E1 through E5.

\subsubsection*{Process}
We conducted semi-structured interviews to capture the perspectives of our target end-users: UX professionals. 
Each interview lasted up to one hour and was conducted via video call. 
The interview was split into two phases.
The first and primary phase was an open ideation session, in which participants reflected on how evaluation is currently conducted in their UX workflows and how they might envision automated systems supporting those efforts.
This phase was deliberately open-ended, allowing participants to express needs and expectations in their own terms. 
To help surface and refine key ideas, participants and facilitators collaboratively used post-it notes on a shared \textit{Miro} board. 
In the second phase, we introduced early mockups and prototypes of potential solutions. 
Participants were invited to give feedback on specific features and discuss what seemed promising or relevant for their own workflows. 
By presenting concrete examples, we aimed to prompt more grounded and reflective responses. 
The full set of guiding questions for the semi-structured interview is provided in Appendix A.


\subsubsection*{Analysis Method}
Each session was recorded and transcribed using automated tools.
We analyzed the interviews using reflexive thematic analysis~\cite{braun_2006_using, braun_2019_reflecting}, working primarily inductively and at a semantic level within a critical-realist frame that treats participants' accounts as reports of genuine workflow constraints.
One author familiarized themselves with the full corpus and generated initial codes across all transcripts, coding openly rather than against a fixed codebook.
The author team then collaboratively constructed candidate themes from these codes, returning iteratively to the coded extracts and full transcripts to develop, review, and define each theme as a pattern of shared meaning organized around a central concept.
Rather than seeking inter-coder agreement, we treated discussion among the authors as a resource for richer interpretation, with our own backgrounds in HCI and UX tooling actively shaping the analysis.
This process produced the final set of themes, which we organized into the design requirements presented in Section~\ref{sec:requirements-list}.


\subsection{Understanding UX Workflows}

To understand user needs, we first established the typical UX workflow our participants engage in. While practices vary across 
contexts, a shared structure emerged that resonates with established models in design research~\cite{designcouncil_2019_doublediamond, brown_2010_designthinking,
hevner_2004_designscience, suh_2025_storyensemble}: four stages of
\textit{Discovery}, \textit{Definition}, \textit{Design}, and
\textit{Delivery}, each paired with evaluation methods ranging from surveys and usability studies to A/B tests and analytics.
Participants agreed that \textit{Discovery}, \textit{Definition}, and late \textit{Delivery} rely on nuanced human perspectives and are least amenable to automation, but expressed cautious optimism about simulation during the \textit{Design} stage, where rapid iteration can benefit from scalable, on-demand feedback. 
As E3 noted, \textit{``At those early stages, I see the most potential. Later, you probably want to hear from users, and that's where I have serious doubts, because we're talking human psychology.''} \looseness=-1


\subsection{User Needs and Challenges}

\revision{Our reflexive thematic analysis produced five themes: \textit{frequent iteration}, \textit{slow cycles}, \textit{isolating and prioritizing findings}, \textit{diversity of perspectives}, and \textit{goal alignment}. 
We present each theme below with illustrative extracts from the participants.}

\revision{
A central theme was the importance of \textit{frequent iteration}. New interfaces cycle through many revisions before release and teams reserve user studies for mature solutions while early prototypes are iterated internally using expert-driven methods such as cognitive
walkthroughs~\cite{lewis_1997_cognitive} and heuristic evaluations~\cite{nielsen_1990_heuristic}. 
E1 described this rhythm as, \textit{``We probably have like 10-20 or even more iterations, not counting minor changes.''} 
With AI-assisted coding shortening build cycles, experts saw clear value in a complementary tool for comparing alternatives at this pace.
}

\revision{
Consequently, \textit{slow cycles} emerged as a recurring pain point.
Live user traffic is scarce and contested across experiments, particularly for niche or newly launched websites. 
Similarly, smaller teams lack resources for frequent studies. 
The most common request was for evaluation that can be generated on demand, rerun after small edits, and structured automatically.
E4 explained, \textit{``I don't think the AI could ever replace a human study. But it could be one more data point, a proxy that we run through to see what it says.''}
}

\revision{
Alongside iteration speed, participants emphasized the challenge of
\textit{isolating and prioritizing problems}. 
The participants remarked on a common two-step practice where quantitative signals, such as metrics, are used to first detect patterns and then qualitative methods, like focus groups, are applied to explain why those patterns occur. 
Participants saw the first step as poorly supported by existing tooling and a natural target for AI-driven approaches. 
E2 illustrated the burden: \textit{``I can get drawn in like 1000 feedbacks saying different things... it is hard to find what to prioritize.''}\looseness=-1
}

\revision{
When pushed on what makes human studies the gold standard, participants pointed to the \textit{diversity of perspectives} that is difficult to access otherwise. 
Teams want feedback mirroring the variety of likely users, yet time, budget, and traffic constraints limit how many perspectives can be consulted. 
E1 emphasized, \textit{``This is definitely a pain point in our
design process, especially because it's really hard to get access to those specific personas we design for.''} 
Even with a well-understood population, and despite helpful priors from historical data, participants found real user behavior difficult to anticipate.
}

\revision{
Finally, participants emphasized \textit{aligning feedback with user goals}.
Evaluations are typically guided by specific intentions, such as completing a task or nudging users toward an outcome – as E2 put it, \textit{``testers rarely explore aimlessly.''} 
Practitioners therefore need to see what a user is trying to achieve at the moment a problem occurs.
Several experts pointed to think-aloud protocols as a model, since they make reasoning explicit and reveal meaningful breakdowns. \looseness=-1
}

\vspace{-0.5em}
\subsection{Description of Requirements}
\label{sec:requirements-list}

\revision{
Taken together, these findings point to a set of design requirements that an automated system should meet in order to effectively support UX professionals, each operationalizing one of the themes:
}

\revision{
\begin{enumerate}[leftmargin=0.7cm]
    \item[R1:] \textbf{Focus on iteration} (\textit{frequent iteration}): Support rapid interface analysis during early-stage design, offering fast, comparable insights across many design permutations.
    \item[R2:] \textbf{Access to perspectives} (\textit{diversity of
    perspectives}): Deliver diverse, persona-aligned feedback that credibly reflects key segments of the target user base and is tunable to different user profiles and contexts.
    \item[R3:] \textbf{Goal-aligned insights} (\textit{goal alignment}): Reveal why issues manifest and where they occur relative to user goals, intentions, and steps along the journey, so practitioners can address root causes rather than symptoms.
    \item[R4:] \textbf{On-demand generation} (\textit{slow cycles}): Enable quick creation of simulation data, rerunning after edits, and automatic structuring of results without manual effort.
    \item[R5:] \textbf{Prioritize findings} (\textit{isolating and prioritizing findings}): Help practitioners navigate large volumes of findings by grouping related issues, estimating severity, and surfacing the most actionable changes first.
\end{enumerate}
}



\begin{figure}[t]
  \centering
  \includegraphics[width=1\linewidth]{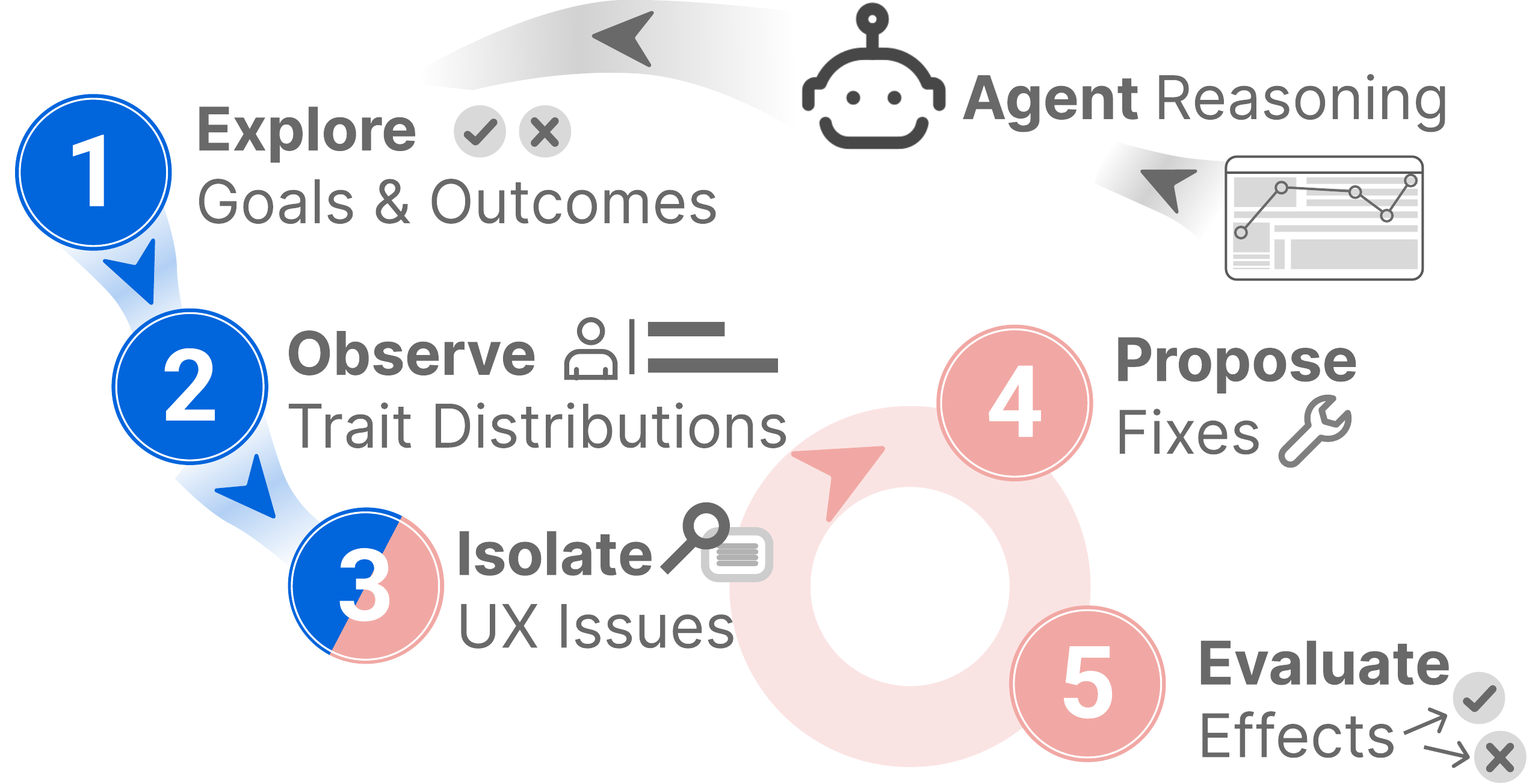}
  \caption{Overview of our workflow combining goal-driven exploration with iterative refinement. Users begin by (1) examining agent goals and outcomes, (2) analyzing behavior across persona traits, and (3) identifying specific UX issues. They can then (4) propose interface changes and (5) assess their effects across diverse agent profiles.
  }
  \label{fig:workflow}
  \Description{
    This process diagram outlines UXCascade's five-step analysis loop.
    Step 1, ''Explore,`` enables users to examine agent goals and their task outcomes.
    Step 2, ``Observe,'' reveals how behavior is distributed across persona traits.
    Step 3, ``Isolate,'' identifies concrete UX issues tied to agent actions and reasoning.
    Step 4, ``Propose,'' allows designers to submit interface fixes.
    Step 5, ``Evaluate,'' simulates how changes affect agent behavior.
    The figure also shows how agent reasoning data supports each step, emphasizing the cyclical and exploratory nature of the workflow.
  }
\end{figure}

\section{Workflow}
\label{sec:workflow}

We illustrate how these requirements can be met using a workflow (\autoref{fig:workflow}) that facilitates a scalable approach to working with simulated user agents.
While persona agents generate detailed action and reasoning traces, the volume and branching complexity of this data make direct analysis challenging. To manage this, our core analysis loop guides practitioners through: \icon{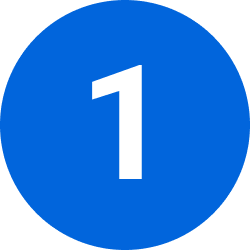} exploring agent goals and task outcomes, \icon{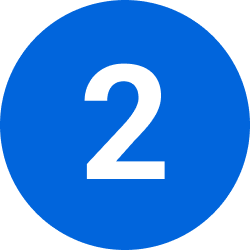} observing trait-based behavior distributions, and \icon{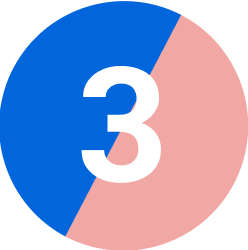} isolating specific usability issues and their underlying causes. 
Practitioners can then \icon{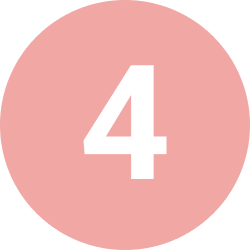} propose design fixes and \icon{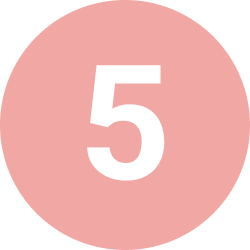} evaluate their impact across diverse personas.

\subsubsection*{\icon{figures/icons/icon1.png}~Explore Goals and Outcomes}
Users begin by reviewing a structured overview of agent goals and task outcomes (R3).
Goals are aggregated from reasoning steps recorded during traversal, grouping low-level behaviors by shared intention.
This top-down entry point lets practitioners scan which objectives were pursued, identify task failures, and prioritize areas aligned with their evaluation focus.\looseness=-1

\subsubsection*{\icon{figures/icons/icon2.png}~Observe Trait Distributions}
Selecting a goal reveals how unsuccessful outcomes are distributed across persona traits (R2).
This view surfaces patterns such as disproportionately low success rates for specific subgroups, helping practitioners isolate which user segments encounter the most friction.

\subsubsection*{\icon{figures/icons/icon3.png}~Isolate Issues and Reasons}
Drilling into a trait or persona reveals a severity-ranked list of concrete UX issues (R5).
Each issue is presented with the agent's think-aloud reasoning, the affected interface element, and a step-level timeline showing how the problem developed.
This view answers not only what went wrong but why, grounding issues in agent goals and interaction context.

\vspace{0.5em}
\noindent These steps organize simulation data into structured overviews that support human-in-the-loop reasoning.
To match the iterative pace of UX practice (R1), the workflow also includes a refinement loop.

\subsubsection*{\icon{figures/icons/icon4.png}~Propose Design Fixes}
Users can propose targeted fixes by editing a specific element or describing the intended change in natural language.
Edits are non-destructive and treated as design hypotheses.\looseness=-1

\subsubsection*{\icon{figures/icons/icon5.png}~Evaluate Effects}
A single simulation step is re-run on the modified snapshot to estimate whether the identified issue has been resolved (R4).
Practitioners can explore how small changes might influence agent behavior without committing to full re-simulation.



\section{\app}
\label{sec:usecases}


We introduce \app, an end-to-end system that generates simulated, persona-grounded usability data and supports our workflow for human-in-the-loop analysis of aggregated insights.
Below, we outline the underlying agent framework along with the components of the user-facing visual interface.

\subsection{Agent Framework}

\app implements three sets of agents to facilitate a scalable UX evaluation workflow (\autoref{fig:agent-framework}). A supplementary technical report with full prompt specifications is provided in \autoref{app:technical-report}.

\subsubsection*{Simulation Agents}

We utilize a custom LLM-based agent built on the \textit{browser-use} framework~\cite{muller_2024_browseruse}, which operates through an iterative perception-decision-action loop. 
At each step, the agent receives a structured state summary derived from an indexed map of interactive DOM elements and annotated screenshots~\cite{yang_2023_setofmark}. 
The agent selects from a task-agnostic action set (e.g., click, scroll, type) based on its internal reasoning. 
To support downstream analysis, the system emits structured event snapshots via an event bus, recording raw HTML, prompt state, and tab-level metadata at the moment of interaction.
\app generates permutations of user-defined characteristics to form diverse persona profiles. 
Throughout the simulation, agents are prompted to align their actions with these assigned traits and justify decisions accordingly. 
Users can further define task-specific goals or custom behavioral directives to reflect specific evaluation scenarios.
\revision{This design ties the simulation to our formative findings: personas instantiate the behavioral axes the study surfaced (R2), and agents pursue explicit task goals (R3).}

\subsubsection*{Annotation Agents}

The \textit{Tagging Agent} generates high-level semantic labels from the reasoning traces of simulated agents to identify behavioral trends across multiple runs. Using "think-aloud" logs as input, the agent identifies cognitive intents such as \textit{exploring}, \textit{searching}, or \textit{deciding}. These categories are informed by HCI research on user goals and interaction patterns. To ensure persona-sensitive representation, the agent iteratively refines its tagging behavior based on a consistency metric that rewards distinctiveness between different personas.
Simultaneously, an \textit{Issue Detector} agent analyzes the same reasoning traces to isolate moments of friction. This agent identifies the specific interface elements involved in a breakdown and assigns a severity level, transforming unstructured logs into a structured database of actionable usability issues.

\subsubsection*{Refinement Agents}

To enable rapid prototyping, the \textit{Editor Agent} supports natural-language editing of HTML snapshots. It interprets user instructions to generate precise DOM-level patches for minor updates or synthesizes entirely new templates for major layout reconfigurations. This allows practitioners to iterate on design hypotheses directly within the tool.
The workflow concludes with a \textit{Preview Agent} that provides automated feedback on proposed changes. It replays critical steps from the original simulation using the modified HTML while maintaining the initial prompt state. The agent evaluates whether the previously observed usability issue still manifests, providing a structured report and a predicted behavior summary to validate the fix in a closed simulation loop.

\begin{figure}[t]
  \centering
  \includegraphics[width=1.0\linewidth]{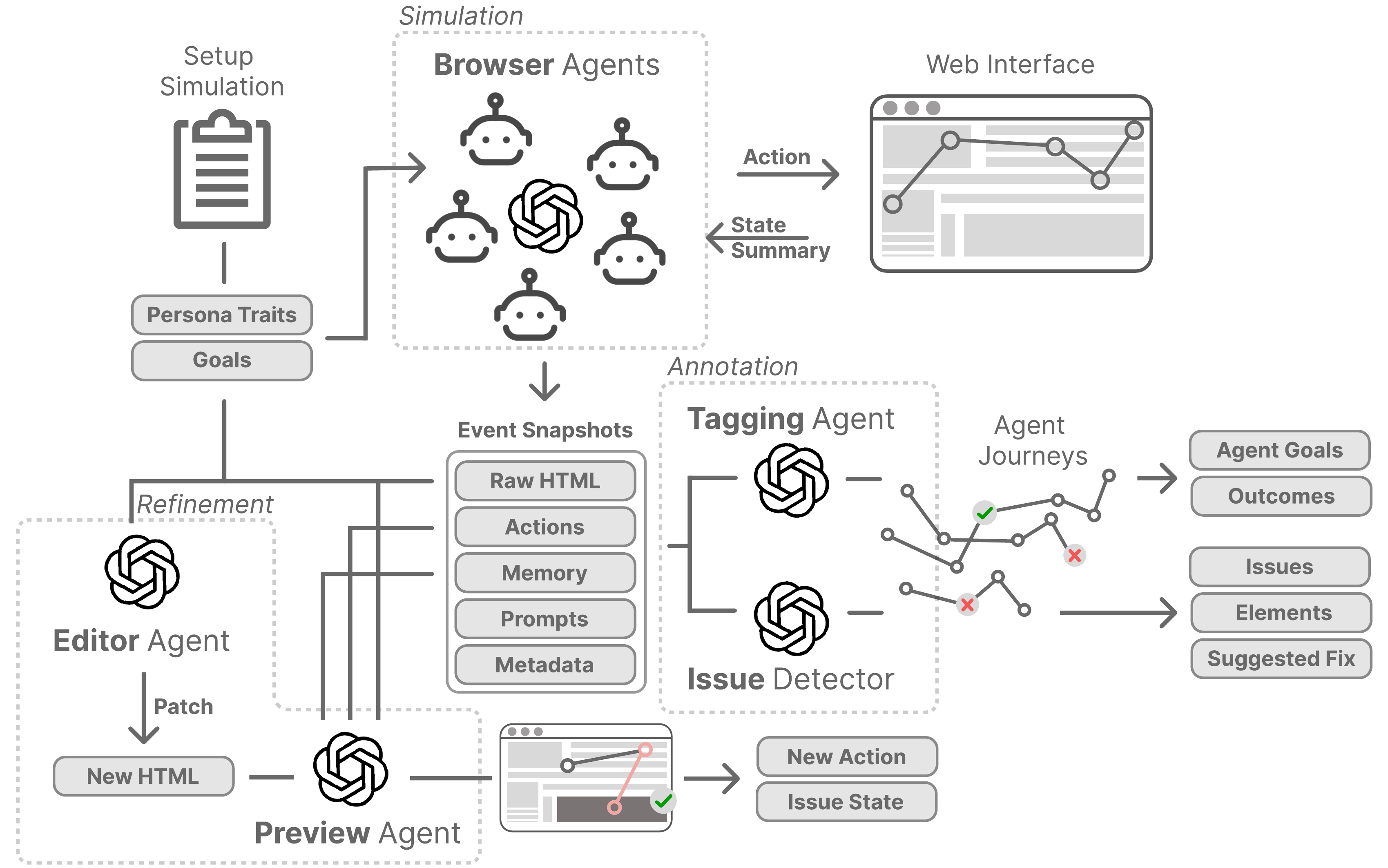}
  \caption{Our agent framework comprising (1) \textit{Simulation Agents} for generating persona-grounded interaction traces, (2) \textit{Annotation Agents} for tagging cognitive intent and detecting usability issues, and (3) \textit{Refinement Agents} for applying and evaluating targeted interface edits.}
  \label{fig:agent-framework}
  \Description{
    This diagram shows the full architecture of the UXCascade system, organized into three key components: Simulation, Annotation, and Refinement.
    The Simulation section depicts browser agents that interact with web interfaces based on persona traits and predefined goals. These agents receive state summaries from the DOM and visuals, then generate actions like clicks or scrolls.
    In the Annotation module, two agents—the Tagging Agent and the Issue Detector—analyze simulation logs to generate cognitive intent labels and detect usability issues with severity assessments.
    The Refinement module includes an Editor Agent that applies user-requested HTML changes, and a Preview Agent that replays relevant steps in a modified version of the site to assess whether the fix addresses the original issue.
    The flowchart outlines how simulation data flows through the system and becomes structured insight for UX evaluation.
  }
  \vspace{-7pt}
\end{figure}

\subsection{Visual Interface}

The interface (\autoref{fig:teaser}) was refined iteratively based on feedback from UX experts from our formative study.  
To better illustrate the tool's functionality and motivate key design choices, we use a running example of a commercial font website, as suggested by one of our study participants.


\begin{figure*}[h]
  \centering
  \includegraphics[width=1\linewidth]{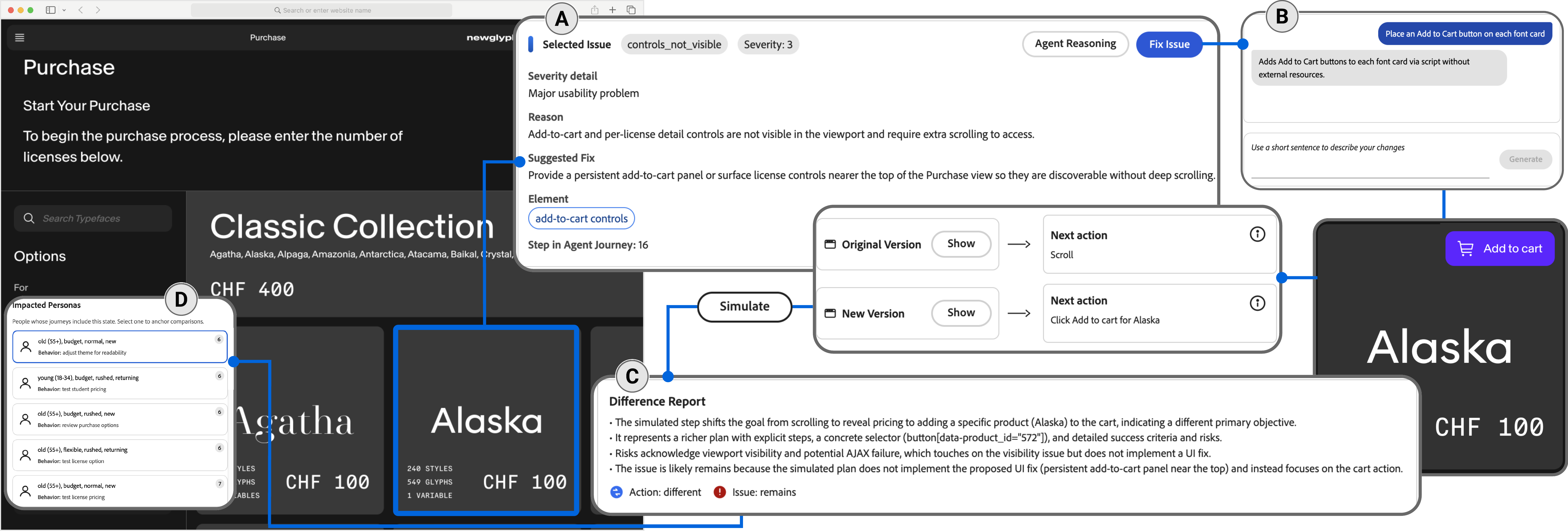}
  \caption{User interaction scenario with the \app interface: (A) isolating issue details, (B) generating a fix via natural language, (C) simulating new agent behavior, and (D) visualizing impacted personas across the agent population.}
  \label{fig:interface-ex}
  \Description{
    This figure illustrates a worked example of the UXCascade workflow applied to a commercial font website.
    Panel A shows an automatically detected issue (``controls not visible'') along with its severity rating, reasoning, and suggested fix.
    Panel B displays the user-generated fix entered via natural language, specifying a design change to place ``Add to Cart'' buttons directly on font cards.
    Panel C compares agent actions before and after the proposed fix, revealing a change from a scrolling action to directly clicking the intended element, highlighting improved discoverability.
    Panel D lists impacted personas whose journeys included the affected interface element, allowing users to assess how changes affect different subgroups.
  }
\end{figure*}

\subsubsection*{Experiment Setup}
Users begin by configuring a simulation experiment that specifies target persona traits and high-level testing goals that help guide agent behavior. 
This is done through a dedicated panel that allows for quick creation of mock participants and formulation of experiment hypotheses. 
Because simulations are computationally lightweight, users are encouraged to explore both common user paths and edge cases without resource constraints. 
Once a simulation is completed, it is listed in the experiment dashboard, where users can initiate downstream analysis sessions.

\subsubsection*{Analysis View}
The analysis panel serves as the central hub for identifying and exploring usability issues (\autoref{fig:teaser}A). 
It first presents a goal-level summary of simulation outcomes-displaying the number of agents attempting each goal, how many encountered issues, and a goal success ratio. 
Users can sort or filter these entries and select a specific goal to investigate further.

Selecting a goal opens the trait analysis view that can be toggled between two complementary modes. 
The trait-centric mode presents issue distributions per individual trait using bar charts, helping users assess how specific characteristics correlate with usability breakdowns. 
Alternatively, the single persona mode highlights the most problematic trait combinations by grouping agents into composite profiles. 
In both modes, users can drill down into a prioritized list of usability issues associated with the selected group, sorted by severity (0-4). \looseness=-1

Each issue reveals a description, the affected interface element, and a proposed fix to guide remediation  (\autoref{fig:interface-ex}A).
Users can also explore additional context: the original think-aloud trace, the annotated screenshot used by the agent, and a timeline of agent decisions. \looseness=-1


\subsubsection*{Agent Journey}
To contextualize simulation data along the temporal dimension, the interface includes agent journey panel rendered as a Sankey diagram.
This panel supports two toggleable modes: one emphasizes page-level navigation, illustrating how agents move through the website structure, while the other aligns agent paths with the step level goals, mirroring the objectives from the analysis view. 
Each node links back into the issue-level analysis.

\subsubsection*{Fix Issues}

The fix issues interface (\autoref{fig:teaser}B) enables lightweight prototyping of design improvements. 
The panel displays a rendered snapshot of the page in question, with the current issue highlighted as a reminder. 
Users can access a toolbar with a simple editor to make surface-level edits by clicking directly on page elements. 
Typical changes include updating labels, modifying links, changing font properties, or removing problematic elements.

For more complex revisions, users can invoke a chat-based interaction with our \textit{refinement agent} to describe desired changes in natural language (\autoref{fig:interface-ex}B). 
The agent applies these changes to the page and provides real-time feedback of the operation.
Users can iterate freely and access a history panel to revert edits as needed. 
Once a satisfactory fix is achieved, they can proceed to evaluation. \looseness=-1

\subsubsection*{Evaluation}

The evaluation panel (\autoref{fig:teaser}C) enables users to compare the pre- and post-edit interfaces and assess whether the proposed changes lead to meaningful improvements. 
The system identifies modified elements and replays the relevant simulation step using the updated snapshot. 
By leveraging the original prompt history, the system estimates whether the agent's next action has changed and whether the original issue has been resolved (\autoref{fig:interface-ex}C).\looseness=-1

To assess broader impact, the system suggests additional personas whose experiences may be affected. 
It does so by reusing existing simulation data to identify agents with the adjacent goals who previously interacted with the modified element (\autoref{fig:interface-ex}D).







\section{User Study}
\label{sec:userstudy} 


\subsection{Study Design}

\subsubsection*{Overview}

We design our user study to reflect a common scenario in the UX development cycle: a first prototype is available, and the practitioner seeks rapid, actionable feedback. 
The focus at this stage is broad, ranging from basic usability problems to engagement-related issues in layout, flow, or content.
We assume that the UX professional has clear design goals in mind and hypothetical client types that they envision interacting with the website. 
This setup mirrors realistic early-stage testing where practitioners aim to validate assumptions and surface hidden issues. 

We evaluate whether \app meets the requirements outlined in \autoref{sec:requirements}, specifically testing whether it helps UX professionals (1) uncover usability issues they would otherwise miss, (2) link issues to user behavior, personas, and interface elements with greater clarity, and (3) support faster iteration by devising fixes.\looseness=-1

\changes{
To do so, we adopt an end-to-end evaluation strategy that examines how well our tool integrates into existing UX workflows. 
The baseline condition consists of a lightweight usability report compiled in advance from human feedback on the same prototype. 
While more sophisticated tools exist for processing such data, we opted for a minimal setup to better reflect the realities of early-stage design teams. 
These teams often operate under time and resource constraints and are likely to benefit most from scalable, simulation-based alternatives such as \app.}

\subsubsection*{Participants}
We recruited eight UX practitioners, each with professional experience (mean=6.5$\pm$4 years) in designing, building, or evaluating web interfaces.
Participants included UX designers, UX researchers, and product engineers who routinely conduct or support interface testing.
For each expert, we recorded their role, years of experience, and domain specialization.
Throughout the study, participants are referred to as P1-P8.

\subsubsection*{Materials and Conditions}
We evaluated a version-controlled custom-built online t-shirt shop (Cascada Tees), designed to reflect interaction patterns found in typical e-commerce websites.
The prototype draws direct inspiration from real-world marketplace interfaces and includes a seeded set of usability and engagement issues, each annotated with a predefined description, severity rating, and potential fix. 
The seeded issues were derived in consultation with experts and based on Web Content Accessibility Guidelines \cite{w3c_2024_wcag}, Nielsen's usability heuristics \cite{nielsen_1994_heuristics}, and recent e-commerce research \cite{baymard_2025_cart}, with each issue mapped to concrete UI elements \revision{(see Appendix B.2)}. 




We compared two within-subjects conditions on the prototype website, with condition order counterbalanced across participants. In (1) \textbf{User Evaluation Report}, participants were provided with a summary compiled from a formative human study (n=10) conducted on the same website. In (2) \textbf{\app}, participants used our tool, which exposed agent-generated reasoning traces, multi-level analysis views, and lightweight editing and evaluation capabilities. \looseness=-1

\revision{
For the \app condition, we instantiated the agents with four binary traits: price sensitivity, time pressure, age cohort, and user type. These are established e-commerce segmentation axes drawn from UX research~\cite{baymard_2025_cart} and mirrored in recent agentic-shopper systems~\cite{mansour_2025_paars, wang_2025_opera}.
Restricting each trait to two contrasting values was a deliberate choice to maximize behavioral contrast while keeping the population manageable ($2^4 \times 2 = 32$ agents).
We regard this schema as one instantiation rather than a general taxonomy.\looseness=-1
}


\subsubsection*{Process} 
Before the main study, we conducted a preliminary asynchronous evaluation to establish a human-generated baseline. 
Ten lay users were recruited to interact with our custom-built t-shirt e-commerce prototype (Cascada Tees). 
Each participant was asked to identify usability and engagement issues using a structured form, reporting the issue description, perceived severity, and how the observed behavior diverged from their expectations. 
These results were compiled into a spreadsheet and anonymized under pseudonyms. 
This corpus simulates the kind of early-stage, low-resource feedback UX practitioners typically collect and served as the input for our baseline condition.

The main part of the study was conducted remotely via Microsoft Teams. 
After obtaining informed consent and explaining the general study goals, participants were then described the overall task of isolating issues on a custom website.  
Throughout the session, we employed a think-aloud protocol to elicit participants' real-time reasoning and reflections across all conditions.

All participants began with a 7-minute free exploration phase, during which they assessed the prototype unaided.
To log findings, participants completed a structured Google Form that captured issue description, severity rating (0-4), impacted persona traits (if applicable), and proposed fixes. 

They were then randomly assigned to one of two counterbalanced conditions \textit{User Evaluation Report} or \app.
For each condition, participants were given 10 minutes to continue identifying issues using the assigned tool. 
\changes{This time window was informed by insights from our formative study, where experts emphasized the need for rapid, iterative reviews during early-stage prototyping. 
In practice, they noted that 10-15 minutes often reflects the realistic time budget available per design iteration or version.}
Before using \app, users viewed a short two-minute demo video outlining key features. 
To mitigate the impact of unfamiliarity, they were encouraged to explore the interface and ask clarifying questions before proceeding.

After completing both conditions, participants filled out the NASA-TLX workload questionnaire. 
This was followed by a 15-minute semi-structured interview to probe perceived usefulness, integration into existing workflows, perceived advantages and limitations, and the utility of specific features.

We captured screen and audio recordings, timestamps, and the full issue lists from each session. 
Interview data was automatically transcribed and analyzed using a thematic analysis approach~\cite{braun_2006_using}. 
The authors coded the data and consolidated themes through discussion. 
Additional materials are included in Appendix~\ref{appendix:protocol}.



\begin{table}[t]
  \centering
  \footnotesize
  \setlength{\tabcolsep}{2.5pt}

\caption{\revision{Distribution of seeded usability issues identified by
  participants, with assigned severity (0-4). Cell
  color indicates the phase in which the issue was found
  (\icon{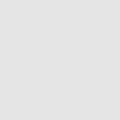} None;
  \icon{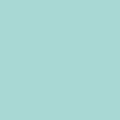} Baseline;
  \icon{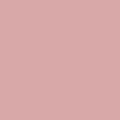} \app{}). The \emph{first/second
  condition} rows indicate the order of the assisted conditions
  (B: Baseline, U: \app{}).}}
    \label{tab:issue-list}
  \renewcommand{\arraystretch}{1.}
  \resizebox{\columnwidth}{!}{%
  \begin{tabular}{@{}l*{8}{c}@{}}
  \toprule
  \textbf{Seeded Issue} & \textbf{P1} & \textbf{P2} & \textbf{P3} & \textbf{P4} & \textbf{P5} & \textbf{P6} & \textbf{P7} & \textbf{P8}\\
  \textit{First condition}  & B & U & B & U & B & U & B & U\\
  \textit{Second condition} & U & B & U & B & U & B & U & B\\
  \midrule
  \multicolumn{9}{@{}l}{\textbf{A: Discovery \& Interaction}}\\
  \addlinespace[1pt]
  A1: Sort label mismatch        & \colBase{4} & \colNo{1}   & \colNo{3}   &             &             & \colBase{4} & \colNo{2}   &             \\
  A2: Price filter unresponsive  & \colBase{4} & \colNo{3}   &             & \colBase{3} &             & \colNo{3}   & \colNo{3}   &             \\
  A3: Filter state not persisted &             &             & \colUX{4}   &             & \colUX{4}   & \colBase{4} &             &             \\
  A4: Inconsistent bundle layout &             &             &             & \colNo{2}   &             &             &             &             \\
  A5: Returns flow hard to find  & \colUX{3}   &             & \colUX{3}   &             &             & \colUX{3}   & \colUX{3}   &             \\
  A6: Vintage filter broken      &             &             &             &             & \colBase{4} &             &             &             \\
  A7: No item-level cart control &             & \colUX{3}   & \colBase{4} & \colNo{3}   & \colUX{4}   & \colUX{3}   &             & \colUX{3}   \\
  \addlinespace[3pt]
  \multicolumn{9}{@{}l}{\textbf{B: Accessibility \& Communication}}\\
  \addlinespace[1pt]
  B1: No size info in bundles    & \colBase{3} &             &             & \colNo{4}   &             &             &             &             \\
  B2: Shipping cost unclear      & \colUX{1}   & \colNo{3}   &             & \colUX{2}   &             & \colUX{2}   & \colBase{2} &             \\
  B3: Progress meter off by one  & \colNo{4}   &             &             & \colNo{3}   &             & \colBase{4} & \colBase{4} &             \\
  B4: Bundle rules unclear       &             &             & \colNo{3}   & \colUX{3}   & \colNo{4}   &             & \colBase{4} & \colBase{3} \\
  B5: No add-to-cart feedback    & \colNo{3}   & \colBase{2} & \colBase{3} &             & \colNo{3}   & \colUX{3}   & \colNo{4}   &             \\
  B6: Add-to-cart looks disabled & \colBase{4} & \colNo{4}   & \colNo{3}   & \colBase{3} & \colBase{4} & \colNo{4}   & \colNo{4}   & \colNo{4}   \\
  B7: No shipping/return info    &             & \colUX{4}   & \colBase{4} & \colUX{2}   &             &             & \colUX{2}   & \colUX{3}   \\
  \addlinespace[3pt]
  \multicolumn{9}{@{}l}{\textbf{C: Visual Design \& Performance}}\\
  \addlinespace[1pt]
  C1: Promo strip hides filters  & \colNo{4}   & \colNo{2}   & \colNo{3}   &             & \colNo{3}   & \colNo{1}   &             & \colNo{3}   \\
  C2: Mixed currency display     &             & \colUX{3}   & \colUX{4}   & \colBase{3} & \colUX{2}   &             &             & \colBase{3} \\
  C3: Redundant search button    &             & \colUX{4}   &             &             &             & \colNo{3}   &             & \colNo{2}   \\
  \bottomrule
  \end{tabular}}
  \Description{
  This table presents the full distribution of usability issues identified by participants (P1-P8) under two conditions: Baseline (human-generated feedback) and UXCascade.
  Issues are listed as rows, each labeled with a code and short name (e.g., A1: Sort label mismatch), organized under three bold category header rows: A (Discovery and Interaction), B (Accessibility and Communication), and C (Visual Design and Performance).
  Columns correspond to participants P1 through P8, with two header rows indicating the order in which each participant experienced the two conditions.
  Each cell is color-coded to indicate the condition under which the issue was identified, and the number in the cell represents the severity score assigned by the participant (0 to 4, higher is more severe).
  This format allows a fine-grained comparison of issue coverage, highlighting overlaps, omissions, and severity perception across methods and participants.
  }
\end{table}

\begin{figure}[t]
    \centering
    \includegraphics[width=\linewidth]{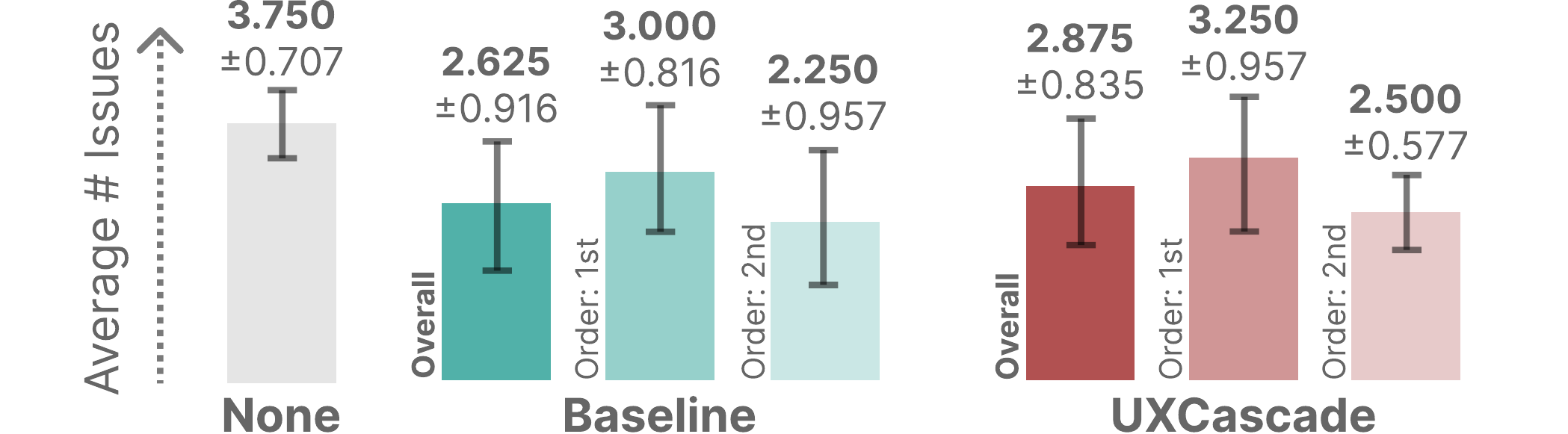}
    \hfill
    \vspace{-2em}
    \caption{Average number of usability issues identified by participants across conditions (error bars: ±SD). Each condition is split by whether it was used first or second, illustrating the order effect.\looseness=-1}
    \Description{
    This bar chart compares the average number of usability issues identified by participants across three feedback conditions: None (no support), Baseline (human-generated feedback), and UXCascade (AI-based simulation).
    Each condition is split into two subconditions representing different participants' task orders.
    Bars show the mean number of issues identified, with error bars representing standard deviations.
    The chart demonstrates that UXCascade identifies a comparable number of issues to the human-generated baseline, indicating potential for practical use in UX workflows.
    }
    \label{fig:aggregate}
    \vspace{-1.5em}
\end{figure}

\subsection{Study Results}
\label{sec:study-results}

Based on our observations and the collected quantitative and qualitative feedback, we outline the key insights gained in the study. 



\subsubsection*{Comparing Conditions}

The issues recorded by participants were compiled and compared to the list of seeded issues embedded in the prototype website. 
Since our study involved UX experts with several years of experience, participants occasionally surfaced novel observations beyond our predefined list. 
While these additional suggestions were documented, we exclude them from the main analysis, as they often involved stylistic preferences or speculative comments about user expectations rather than concrete usability problems.
This also made them difficult to evaluate systematically.\looseness=-1

The complete set of results for each condition is summarized in~\autoref{tab:issue-list}. 
We indicate which seeded issues were identified by (1) no condition, (2) the baseline condition, and (3) the \app condition.
For each issue, we also report the severity ratings assigned by participants. The aggregation of the results is shown in~\autoref{fig:aggregate}.
\revision{
Given the small sample and the order effect described below, we treat the comparisons between conditions as exploratory.
A Wilcoxon signed-rank test on per-participant issue counts shows no significant difference between conditions (W=6, p=.813).
}

Overall, the highest number of issues (3.750$\pm$0.707) was identified during the initial onboarding step, where participants explored the website without tools. 
This was followed by the \app condition (2.875$\pm$0.835) and the baseline condition (2.625$\pm$0.916), which yielded the fewest issues. 
This pattern may partially be explained by order effects, as participants examined the same website multiple times. 
The most obvious issues were often identified early, making later phases more challenging. 
For instance, issues A1 and A2 — both related to filtering behavior, were frequently identified during the first phase.
This trend is further illustrated in~\autoref{fig:aggregate}, where both assisted conditions resulted in fewer issues when presented as the second condition. 
However, the marginal advantage of \app remains evident when stratifying by condition order. 
In both configurations, participants identified slightly more issues using our system than with the baseline.\looseness=-1



\begin{figure}[t]
  \centering
  \includegraphics[width=1.0\linewidth]{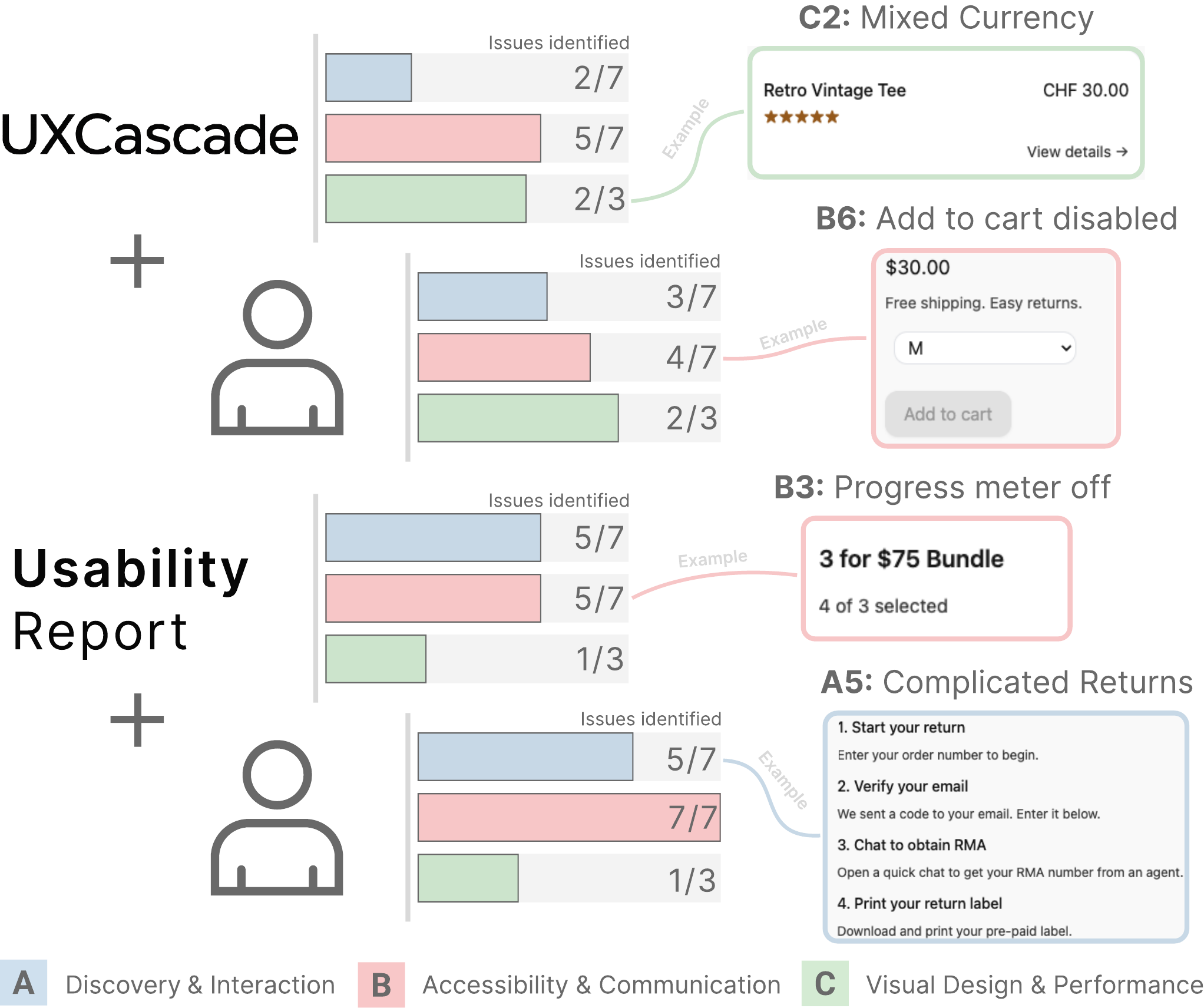}
  \vspace{-2em}
  \caption{\revision{Seeded issues covered per category (A–C) for \app and the Usability Report (Baseline). Comparison of each tool's own output (top) vs. participants using it \icon{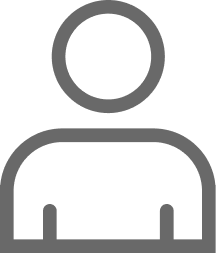} with one example issue per block.}}
  \label{fig:issue-categories}
  \Description{
  Four groups of bar charts comparing issue coverage. For UXCascade: bars showing issues surfaced by agents per category (2 of 7, 5 of 7, 2 of 3) and issues identified by participants using it (3 of 7, 4 of 7, 2 of 3). For the Usability Report: issues contained in the report (5 of 7, 5 of 7, 1 of 3) and issues identified by participants using it (5 of 7, 7 of 7, 1 of 3). Each block shows one representative example issue, such as mixed currency display or the disabled-looking add-to-cart button.
  }
  \vspace{-1.5em}
\end{figure}

\revision{
\subsubsection*{Agent Coverage of Seeded Issues}
To assess the validity of the agent pipeline itself, we mapped every
agent-surfaced issue to the seeded ground-truth set (\autoref{fig:issue-categories}).
Because an agent's suggestion acts primarily as a pointer rather than a verdict, exact matches are not required.
As such, we used permissive keyword and pattern rules to construct the mapping. 
At least one agent directly surfaced 9 of the 17 seeded issues, consistently identifying the returns, currency, shipping, and add-to-cart problems, while the static report contained 11 of 17. 
Notably, the two sources exhibit complementary coverage profiles: the human report was strongest on discovery and interaction issues (5/7 vs.\ 2/7 for the agents), whereas the agents matched or exceeded it on communication and visual consistency issues (e.g., 2/3 vs.\ 1/3 in category C), and each surfaced seeded issues the other missed entirely (e.g., the off-by-one progress meter appeared only in the report, the convoluted returns flow only in agent output). 
This supports positioning simulation as a complement to human feedback rather than a substitute. 

We further regard agent coverage as a lower bound on utility: an agent that misses the exact issue can still point the practitioner to the affected region, where they surface the fault themselves. 
For the same reason, a clean false-positive rate is difficult to define, as agents also surface valid out-of-set issues that the workflow treats as candidates for consolidation rather than detector errors.
Agent-assigned severity tracked expert ratings closely. For example, the mixed-currency issue (C2) was flagged by 32 agents with a mean severity of 3.1 against the mean rating of 3.0 from our expert participants. 
The full per-issue breakdown is provided in Appendix~\ref{appendix:coverage}.
}

\subsubsection*{Applying our Workflow}

Observing how participants used both \app and the baseline condition provided valuable insights into the effectiveness of our proposed workflow. 

The core analysis loop that begins with goals, moves through traits, and ends with issues was found intuitive by most participants. 
P1 shared that \textit{``It collects all the most important highlights... you don't have to dig into the information,''} and added, \textit{``The breakdown by goals and traits makes it much easier to focus compared to the Baseline condition.''}
The persona-driven design also resonated strongly. 
P8 mentioned, \textit{``I liked that you can go by persona and see which ones fail where.''} P1 noted, \textit{``I usually work with personas already, so this just felt like an extension of that.''}

Several participants appreciated the system's ability to reveal patterns that might otherwise go unnoticed. 
P3 said, \textit{``I could imagine this uncovering some patterns you wouldn't necessarily find just manually going through a page.''} 
P4 added, \textit{``It's doing what a checklist or report can't do... it's allowing active scanning.''}

While the editing and simulation features were not central to the assigned task, multiple participants experimented with them and recognized their value. 
P7 commented, \textit{``For my personal workflow, I rely heavily on those tiny little fixes and being able to directly go in and fix something will be immensely helpful.''} 
P6 noted its utility for designers in time-constrained environments \textit{``You could make a bunch of tiny changes without involving dev time.''} 
P4 reflected on using the functionality to simulate a step with the new design, saying, \textit{``Editing the button label and then seeing the next action change was a really cool moment.''}

\subsubsection*{User Perception towards \app}

\begin{figure}[t]
  \centering
  \includegraphics[width=1.0\linewidth]{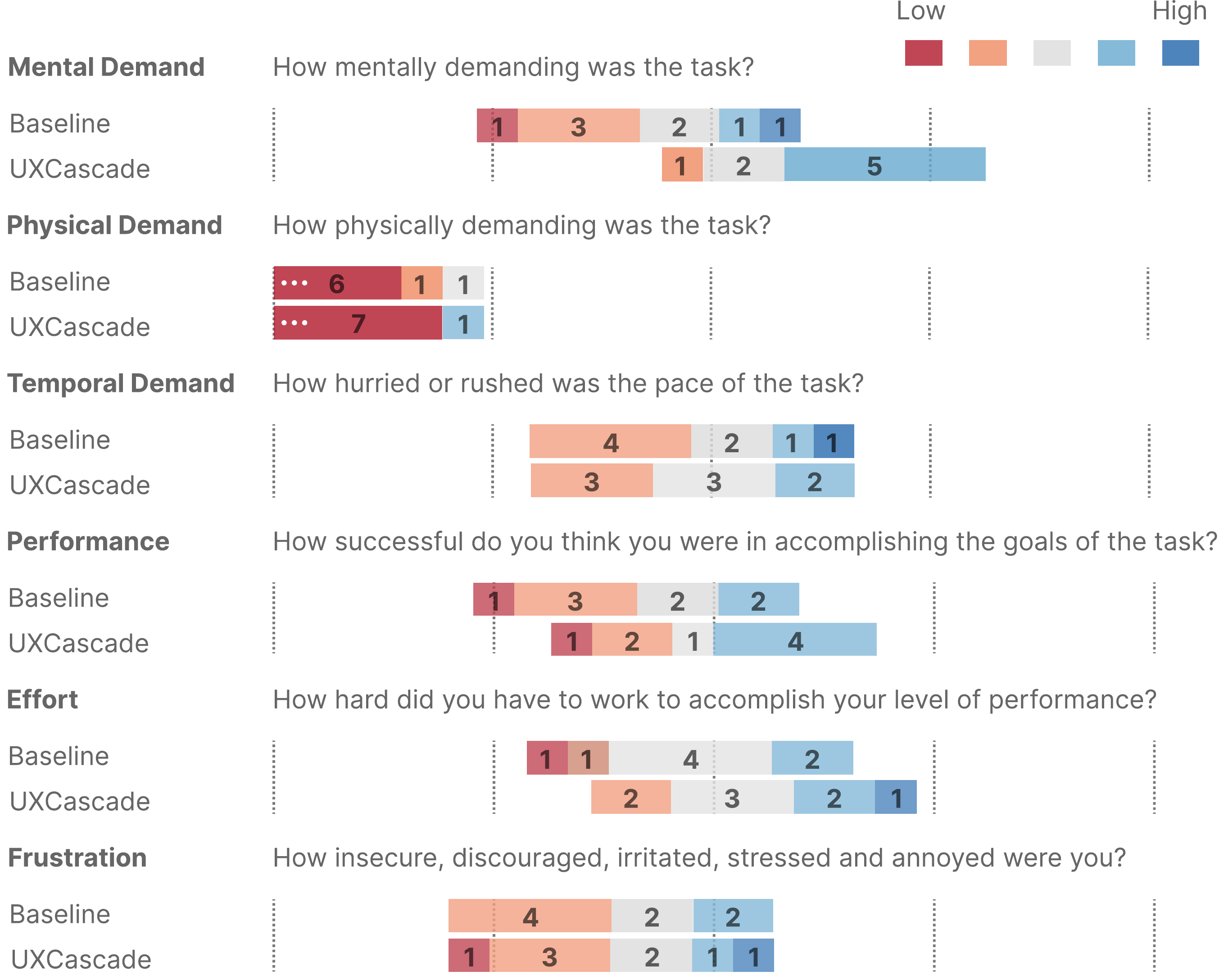}
  \caption{NASA-TLX responses (n=8) for each workload dimension under the Baseline and \app conditions.}
  \label{fig:nasa-tlx}
  \Description{
    This figure presents a visual breakdown of participants' self-reported workload across seven NASA-TLX dimensions: Mental Demand, Physical Demand, Temporal Demand, Performance, Effort, and Frustration.
    Stacked bars compare the Baseline and UXCascade conditions for each metric, color-coded from red (low) to blue (high).
    The visualization reveals how participants perceived the cognitive and emotional burden of completing tasks with each system, as well as their satisfaction with performance.
    UXCascade shows slightly higher perceived success and less frustration, while physical demand remained constant across conditions.
  }
  \vspace{-13pt}
\end{figure}

We used the NASA Task Load Index (NASA-TLX) to assess subjective workload across six dimensions. 
Following established practice in HCI user studies, we omit the pairwise weighting step and report both the unweighted average and individual subscale scores (\autoref{fig:nasa-tlx}).

\revision{Overall, workload was comparable across conditions: no NASA-TLX dimension  differed significantly, and we interpret the subscale patterns below as exploratory.
The two dimensions where \app was rated higher relate to interface richness.
Mental demand was rated slightly higher for \app (3.50$\pm$0.76) than the Baseline (2.75$\pm$0.89; Wilcoxon signed-rank, $p=.328$), and participants likewise reported slightly more effort (3.25$\pm$1.19 vs. 2.875$\pm$1.06). This aligns with qualitative feedback indicating that the system has a steeper learning curve and that navigating the interface requires acclimatization.
Several participants described the raw baseline issue list as overwhelming and the structured presentation in \app as easier to navigate, suggesting the added demand reflects the richer interface rather than the analysis itself.
Conversely, participants perceived \app as better supporting performance and task success (3.00$\pm$0.71 vs.\ 2.625$\pm$0.99). \looseness=-1}

\subsubsection*{\app in Practice}

In addition to completing the assigned task, participants were asked to reflect on how they might envision using the proposed tool in their own workflows. 
Several users noted that \app aligns well with early-stage UX iteration practices. 
P3 commented, \textit{``This fits well when you just need fast feedback without launching a full study.''} 
This sentiment was echoed by P2 and P4, who saw the biggest impact either at the beginning of the design cycle as a first pass before conducting other usability tests, or as a final sanity check to ensure that nothing was missed.
P2 also remarked jokingly, \textit{``I already put my design into ChatGPT to check if I missed something.''}
P4 summarized the tool's fundamental value by noting that, given the current lack of comparable solutions, \textit{``it's better than simply having nothing.''} 
P2 further added, \textit{``Sometimes I just need ideas to get started and looking.''} \looseness=-1

Regarding generalization, many participants expressed confidence that the tool would be even more valuable in complex real-world scenarios. 
In particular, P7 envisioned significant potential if the change evaluation step could be made more robust: \textit{``I could then test many different options and see if the result changes.''} 
They referred to use cases such as conversion optimization, where even small adjustments (e.g., banner size) could impact user behavior. 



\subsubsection*{Trust and Realism of Simulations}

A recurring topic in the interviews was the degree of trust participants placed in the simulated agents' ability to identify issues. 
While some users expressed enthusiasm for the system, they also raised concerns about the realism of agent behavior. 
For example, P7 remarked, \textit{``To be honest, I wouldn't 100\% believe in the simulations... whether it is actually reflective of real users' behavior''}, though they added, \textit{``still just having that feature, I think it's quite awesome.''} 
Others were more critical about how well the agents can embody personas with one participant noting, \textit{``To make the AI very persona-aligned is very difficult actually... in my experience the AI is still not the best at perceptional things.''}\looseness=-1

Participants also drew direct comparisons between the issues surfaced in the baseline and those identified in \app. 
P4 appreciated the nuance and emotional tone present in human-generated feedback, highlighting a specific example: \textit{``this promo code feels like a scam''}, which they felt captured user frustration in a way that simulations did not. 
In contrast, agent responses were seen as more methodical and structured. 
P6 echoed this, observing that the system failed to convey the urgency of a problem, whereas the emotional component in the human comments made their impact more immediately clear.

Conversely, several participants were observed to trust \app more readily, often accepting its output at face value while double-checking the human feedback in the baseline. 
The structured and systematic presentation of agent findings may have contributed to a higher perceived sense of reliability.
In general, not all feedback was critical of the agents either. 
P1 was notably optimistic: \textit{``AI is much better in analyzing it (the webpage) and providing a direct fix... but it's not 100\% accurate, it's just a data-based assumption, but it feels realistic.''} They also cautioned that identifying \textit{``more issues''} does not necessarily equate to \textit{``better issues.''}

\subsubsection*{Desired Improvements}
Participants suggested three refinements. P2 wanted a more flexible,
bottom-up exploration loop: \textit{``I want to see all the issues firstand then apply filters.''} P6 and P7 asked for stronger visual anchoring, finding the annotated screenshots insufficient for issues spanning multiple elements. 
Finally, P5 noted an initial learning curve as
\textit{``it took me a moment to understand what I'm looking at.''} 
We synthesize these into design implications in Section~\ref{sec:discussion}.\looseness=-1

\section{Discussion}
\label{sec:discussion} 


Our user study suggests that \app achieves performance on par with human feedback, while also offering several unique strengths. 
\revision{In particular, the system structures analysis along familiar and interpretable dimensions such as persona traits and agent goals, which participants described as easier to navigate than an unstructured issue list.}
This structure encourages users to tailor their analysis to individual preferences while benefiting from an adaptive balance between control and automation.
Although the richness of features introduces a steeper learning curve, participants acknowledged the utility of each component.

The positive qualitative feedback from both the formative and user study further underscores the potential of our tool. 
Even in scenarios where the simulated evaluation underperforms compared to traditional methods, the low cost and minimal risk of adding it to an existing workflow presents strong practical value. 
In such cases, \app can serve as a complementary step, reserving human time for deeper insights related to emotion, tone, or aesthetics—areas that are difficult for current LLMs to assess. 

\changes{However, the introduction of any automated tool is not without risk. 
Overreliance on such systems could lead to professional complacency, with practitioners delegating critical evaluation tasks prematurely. 
This concern is particularly salient given the increasing automation of the design and development pipeline, leaving usability testing as one of the final human-centered checkpoints. 
To mitigate this, our approach centers the analyst, ensuring transparency around model limitations and positioning the system as an informative, not conclusive, aid in the evaluation process. 
Encouragingly, the qualitative feedback from participants indicated a healthy skepticism toward AI-generated outputs, with professionals expressing a desire to verify suggested issues before acting on them.} \looseness=-1

Our workflow, which guides users from goals to traits to issues, closely mirrors the problem-solving strategies employed by UX experts themselves. 
Several participants noted that these steps reflect how they naturally approach early-stage evaluations. 
The refinement loop, which enables lightweight edits followed by simulation-based feedback, was also well received. 
Participants saw this feature as aligning with the iterative and fast-paced nature of UX development. 
Given the flexibility of our approach, we envision it can be readily adapted to other systems, beyond the scope of this study.

\revision{
Finally, participants' suggested improvements generalize into design implications for tools that structure simulated-user output. 
First, \textit{support bi-directional exploration}: our goal-first hierarchy suited large issue volumes, but practitioners working on simpler sites wanted to survey the full issue space and filter down. 
Analysis workflows should let users enter top-down from goals or bottom-up from issues, rather than enforcing a single path. 
Second, \textit{anchor findings visually}: textual issue descriptions with highlighted screenshots were not always sufficient,
particularly when multiple elements interact. 
Agent reasoning should be tied to its concrete on-page location, for example through overlays rather than detached snapshots. 
Third, \textit{disclose complexity progressively}: the wide array of features justify itself on complex sites but felt excessive in simpler settings. 
Therefore, tools could default to a streamlined base version, revealing richer functionality as the issue space grows.
}


\vspace{-0.5em}
\section{Limitations and Future Work}
Our results demonstrate the potential usefulness of our workflow for utilizing simulated agents for usability testing. 
However, we acknowledge several limitations and avenues for future work. \looseness=-1 


\app has demonstrated the ability to generate plausible behavior sequences, but its alignment with real human behavior can be improved. 
Pilot comparisons between agent and human user behavior reaffirmed our hypothesis by revealing significantly lower variability in clicks and elements explored. 
Because personas guide the agents' planning and reasoning, this also highlights a limitation in how we currently define personas. 
They are primarily specified through a small set of behavioral traits and demographic attributes.
Prior work has shown that human behavior cannot be adequately captured by a few keywords and at best they act as over simplifications. 
Instead, LLMs require richer and more contextually grounded input to mimic real decision-making. 
Additionally, we observed that the browser agent consistently struggles with visual perception and reasoning. 
It has difficulty recognizing subtle visual cues and is not well suited for providing feedback on aesthetic aspects such as visual appeal or the naturalness of layout.
Therefore human alignment will require advances on both fronts: more expressive persona conditioning and improved perception of visual context.\looseness=-1

\revision{Our evaluation also bounds our claims: the parity result is relative to a deliberately minimal baseline, and whether benefits stem from agent simulation, structured presentation, or both remains open. 
Future work should include a simulation-vs-presentation ablation, comparison against LLM-based systems such as UXAgent~\cite{lu_2025_uxagent}, and larger studies on organically complex websites.}

In the current implementation, running agent simulations remains decoupled from the analysis view and cannot yet be executed at scale in real time.
Similarly, the interface for evaluating UX fixes is limited to re-simulating a single step for a single agent. 
This is a meaningful constraint: interface changes rarely operate in isolation, and what improves the experience for one persona may introduce friction for another. 
The current system partially addresses this by surfacing other agents that encountered the same snapshot through the journey view, but it does not quantify a fix's impact on their downstream behavior.
\revision{
However, because the system already tracks which agents traverse each state (\autoref{fig:interface-ex}D), a natural next step is to extend the browser agent to multi-step simulation and re-simulate only the personas a change affects, rather than the full pool. 
This would reframe the evaluation loop as a UX regression test: after a fix, the system replays the affected personas across their journeys and diffs the resulting behavior distribution to confirm the issue is resolved, surface newly introduced ones, and help practitioners identify the most robust solution across diverse user profiles.
}

Finally, enabling direct comparison of multiple interface versions, in which agents are exposed to design alternatives of the same page, would strengthen confidence in surfaced issues and match the comparative exploration our participants described. 
We envision the core analysis flow extending to multi-version evaluation with minimal adaptation.


\section{Conclusion}
We introduced \app, a system that brings simulated persona agents into early-stage UX evaluation. Our workflow enables practitioners to uncover usability issues by navigating agent goals, reasoning traces, and persona traits. 
Similarly, by integrating lightweight edit capabilities and a structured evaluation loop, the system supports actionable refinement of target interfaces. 
This combination of simulation, annotation, and iteration extends the reach of usability testing beyond traditional constraints and integrates seamlessly with existing workflows. 
Through a study with UX experts, we demonstrate how \app can help reveal issues that may otherwise be missed.
We view this work as a step towards more intelligent tools that empower practitioners to evaluate and iterate with greater speed, clarity, and control.

\newpage

\bibliographystyle{ACM-Reference-Format}
\bibliography{main}



\clearpage
\appendix

\section{Formative Study}
\label{appendix:formative}

We conducted semi-structured interviews in two parts: the first to understand participants' general UX testing workflows and expectations for AI-assisted tooling, and the second to reflect on the features of a potential system and initial prototypes. 
The overview of the guiding questions are summarized below.

\noindent
\begingroup
\setlength{\parindent}{0pt}
\setlength{\tabcolsep}{0pt}

\renewcommand{\arraystretch}{1.25}
\begin{tabularx}{\columnwidth}{@{}p{0.28\columnwidth}X@{}}

\multicolumn{2}{@{}l}{\textbf{Part 1: UX Testing Practices and Expectations (40min)}} \\

\parbox[t]{0.33\columnwidth}{1. UX Testing\\Practices} &
\textit{How do you typically evaluate or test UX/UI designs and features? For example, do you use internal reviews, heuristics, usability tests, user feedback, or analytics data?} \\

\parbox[t]{0.33\columnwidth}{2. Usability\\Testing Methods} &
\textit{How is usability testing usually conducted in your workflow? What kinds of data do you collect (e.g., click paths, survey results, interviews), and how is it typically analyzed?} \\

\parbox[t]{0.33\columnwidth}{3. Imagining AI\\Support} &
\textit{Can you imagine ways in which AI could assist your UX iteration process? What kinds of insights or automations would be most valuable?} \\

\parbox[t]{0.33 \columnwidth}{4. Design\\Requirements} &
\textit{What are the most important features or requirements for such a tool to be practical in your work? What would make it trustworthy and usable?} \\

\\[-0.75em]
\multicolumn{2}{@{}l}{\textbf{Part 2: Prototype Feedback (20min)}} \\

\parbox[t]{0.33\columnwidth}{5. Workflow\\Alignment} &
\textit{How well do you think the prototypes we showed match how you currently approach UX evaluation? Which parts of the proposed logic align with your process, and which do not?} \\

\parbox[t]{0.33\columnwidth}{6. Feature\\Usefulness} &
\textit{Which features did you find most useful or promising? Were there any that seemed less relevant? How might each feature benefit your work compared to alternatives?} \\
\end{tabularx}
\endgroup


\section{User Study}

In this section, we provide additional details on the user study described in~\autoref{sec:userstudy}, including the full study protocol and the complete list of seeded issues used in the custom website.

\subsection{Protocol details}
\label{appendix:protocol}

The following protocol was used for conducting the user study:
\begingroup
\setlength{\parindent}{0pt}
\setlength{\tabcolsep}{0pt}
\begin{tabularx}{\columnwidth}{@{}p{0.26\columnwidth}X@{}}
\textbf{Pre-Study} &
Conduct asynchronous usability tests with 5-10 lay participants on the target website. 
Feedback is compiled into a raw baseline report summarizing usability and engagement issues. \\[0.75em]

\parbox[t]{0.33\columnwidth}{\textbf{Introduction\\ (5 min)}} &
Welcome participants and collect informed consent. 
Background info is gathered, including role, years of experience, and typical evaluation flows. \\[0.75em]

\parbox[t]{0.33\columnwidth}{\textbf{Onboarding\\ (7 min)}} &
Participants interact with the website independently. 
They identify UX issues, assign severity, link problems to personas/goals, and suggest fixes. \\[0.75em]
\end{tabularx}
\endgroup

\begingroup
\setlength{\parindent}{0pt}
\setlength{\tabcolsep}{0pt}
\begin{tabularx}{\columnwidth}{@{}p{0.26\columnwidth}X@{}}

\parbox[t]{0.33\columnwidth}{\textbf{Condition 1\\ (10 min)}} &
Participants are randomly assigned to either the baseline report or \app.
They continue augmenting the initial usability issue list. 
Those in the \app condition first watch a short demo video and are given time to familiarize themselves with the interface. 
\\[0.75em]

\parbox[t]{0.33\columnwidth}{\textbf{Condition 2\\ (10 min)}} &
Participants switch to the alternate condition. 
They finalize their list of identified issues. \\[0.75em]

\parbox[t]{0.33\columnwidth}{\textbf{Questionnaire\\ (5 min)}} &
Participants complete the NASA-TLX to assess workload and cognitive demand across both conditions. \\[0.75em]

\parbox[t]{0.33\columnwidth}{\textbf{Discussion\\ (10 min)}} &
A short semi-structured interview collects feedback on usefulness, workflow fit, feature utility, and improvement ideas. 
Optionally, \app is shown on a website of the participant's choosing to highlight real world applicability. \\
\end{tabularx}
\endgroup

\vspace{2em}
\noindent When initializing the agents for \app, we generated diverse personas by permuting the following trait dimensions.
Two agents were created for each combination to ensure further diversity.

\begin{itemize}[label=--, noitemsep]
    \item \textit{Price Sensitivity (PS)}: budget, flexible
    \item \textit{Time Pressure (TP)}: rushed, normal
    \item \textit{Age Cohort (AC)}: 18-34 , 55+
    \item \textit{User Type (UT)}: new, returning 
\end{itemize}

\revision{The four chosen traits are e-commerce segmentation axes drawn from UX research~\cite{baymard_2025_cart} and recent agentic-shopper systems with similar dimensions~\cite{mansour_2025_paars, wang_2025_opera}. 
The binary values were a deliberate design decision to maximize behavioral contrast while keeping the population manageable ($2^4 \times 2$ agents).}

Similarly, we assign the following high-level goals to guide the agent behavior. 
For consistency the same success goals are also surfaced to the participants in our pre-study used to compile the baseline conditions. 
This allows them to also guide their exploration along these tasks. 

\begin{itemize}[label=--, noitemsep]
  \item Save as much as possible using bundles or coupons
  \item Find a specific tee under a price target, compare options, and select one using filters
  \item Explore size options and availability, assess clarity and helpfulness of information
  \item Review cost summary, ensuring clarity of subtotal, discounts, and shipping
\end{itemize}

\vspace{1em}
\begin{table*}[htbp]
\centering
\caption{Seeded usability issues embedded in the evaluation website, including severity ratings and recommended fixes.}
\label{tab:seeded-issues}
\resizebox{\textwidth}{!}{%
\renewcommand{\arraystretch}{1}
\begin{tabular}{cllcp{7.2cm}p{6.2cm}}
\toprule
\textbf{Tag} & \textbf{Area} & \textbf{Issue} & \textbf{Severity} & \textbf{Description} & \textbf{Suggested Fix} \\
\midrule
A1 & Product List & Sort label mismatch & 3 & ``Low to High'' selected, but results show high to low & Fix comparator logic to match selected sort label \\
A2 & Product List & Price filter unresponsive & 2 & Slider moves but does not affect results & Wire slider to filtering logic and trigger requery \\
A3 & Product List & State not persisted & 2 & Filters and sorting reset on page reload or back nav & Sync state to URL and restore on load \\
A4 & Shop vs Bundles & Inconsistent layout and filters & 3 & Bundle view lacks filters and there is layout inconsistency & Unify filters and structure across pages \\
A5 & Returns & Returns flow hard to find & 2 & Multi-step process contradicts ``Easy Returns'' claim & Provide a simplified, consolidated return form \\
A6 & Product List & Vintage filter broken & 1 & Vintage filter does not return results & Fix filter query logic \\
A7 & Cart & No item-level control & 2 & ``Remove'' clears all items at once & Enable single-item removal and links to products\\
\midrule
B1 & Bundles & No size info shown & 3 & Users can't see selected sizes before checkout & Require selection and show bundle summary \\
B2 & Checkout Summary & Shipping cost unclear & 2 & Shipping only appears at final step & Show shipping line and estimate earlier \\
B3 & Bundles & Progress meter off by one & 4 & Meter shows 2/3 when all 3 items are picked & Fix meter logic to reflect correct count \\
B4 & Bundles & Bundle rules unclear & 3 & Discount only appears late, no guidance during build & Preview discount and guide bundle progress \\
B5 & Product Page & No feedback on add to cart & 4 & Only feedback is cart count increment & Add ``Added to cart'' confirmation message \\
B6 & Product Page & Add to cart looks disabled & 4 & Button appears inactive but still works & Update button style and fix accessibility \\
B7 & Policies & No shipping or return info & 2 & Missing contact and policy details near CTA & Add policy summary and trust\\
\midrule
C1 & Product Page & Promo strip hides filters & 4 & Banner reduces access to controls above the fold & Shrink banner, adjust layout spacing \\
C2 & Product Page & Mixed currency display & 2 & Some prices in CHF, others in USD & Normalize currency or indicate user locale \\
C3 & Search Bar & Redundant search button & 1 & Button duplicates enter key functionality & Remove or repurpose button \\
\bottomrule
\end{tabular}%
\Description{
    Table of all 17 seeded usability issues embedded in the evaluation website. Each row gives the issue tag (A1 through C3), the affected site area, a short issue name, its severity rating from 1 to 4, a one-line description of the problem, and a suggested fix. Issues span three categories: Discovery and Interaction, Accessibility and Communication, and Visual Design and Performance. Examples include a sort label mismatch on the product list, a bundle progress meter that is off by one, and mixed currency display on the product page.
}
}
\end{table*}

\subsection{Full List of Issues}

Below we present the full set of seeded usability issues embedded into our custom e-commerce website.
These issues were developed in collaboration with UX experts from our formative study and span a range of realistic interface problems.
To facilitate structured analysis, we grouped them into three overarching categories: (A) \textit{Discovery and Interaction}, (B) \textit{Accessibility and Communication}, and (C) \textit{Layout, Visual Design, and Performance}.
The complete list of issues is shown in~\autoref{tab:seeded-issues}.

\subsection{Per-Issue Agent Coverage}
\label{appendix:coverage}
\revision{To complement the aggregate coverage reported in
Section~\ref{sec:study-results}, Table~\ref{tab:agent-coverage} provides the full per-issue breakdown. For each seeded issue, we report the number of agents (of 32) whose output mapped to it under our permissive keyword and pattern rules, the mean severity those agents assigned, and the mean severity assigned by our expert participants (P1--P8, across all study phases). 
Agents additionally reported issues outside the seeded set.
This is consistent with our workflow and we treat these as consolidation candidates rather than detector errors and exclude them from this table.
Agent and participant severities agree within ${\sim}0.5$ for most covered issues.
The notable divergence is B2 (agents 3.1 vs.\ participants 2.0), where agents consistently over-weighted missing shipping information. 
We note that participant ratings from the \app phase may be partially
anchored by displayed agent severities, so this agreement should be read as consistency rather than independent validation.
}

\begin{table}[h]
  \centering
  \footnotesize
  \setlength{\tabcolsep}{4pt}
  \caption{Per-issue agent coverage: number of agents (of 32) whose output mapped to each seeded issue, with mean agent-assigned severity against the mean severity assigned by expert participants. Issues marked with --- were not surfaced by any agent.}
  \label{tab:agent-coverage}
  \renewcommand{\arraystretch}{1.0}
  \begin{tabular}{@{}ll@{\hspace{6pt}}ccc@{}}
  \toprule
  \textbf{ID} & \textbf{Seeded issue} & \textbf{Agents} & \textbf{Agent\ sev.} & \textbf{Participant.\ sev.} \\
  \midrule
  A1 & Sort label mismatch        & 1   & 2.0 & 2.6 \\
  A2 & Price filter unresponsive  & --- & --- & 3.2 \\
  A3 & Filter state not persisted & --- & --- & 4.0 \\
  A4 & Inconsistent bundle layout & --- & --- & 2.0 \\
  A5 & Returns flow hard to find  & 4   & 3.0 & 3.0 \\
  A6 & Vintage filter broken      & --- & --- & 4.0 \\
  A7 & No item-level cart control & --- & --- & 3.3 \\
  \midrule
  B1 & No size info in bundles    & 12  & 3.0 & 3.5 \\
  B2 & Shipping cost unclear      & 32  & 3.1 & 2.0 \\
  B3 & Progress meter off by one  & --- & --- & 3.8 \\
  B4 & Bundle rules unclear       & --- & --- & 3.4 \\
  B5 & No add-to-cart feedback    & 8   & 3.0 & 3.0 \\
  B6 & Add-to-cart looks disabled & 8   & 3.4 & 3.8 \\
  B7 & No shipping/return info    & 32  & 2.8 & 3.0 \\
  \midrule
  C1 & Promo strip hides filters  & 9   & 2.5 & 2.7 \\
  C2 & Mixed currency display     & 32  & 3.1 & 3.0 \\
  C3 & Redundant search button    & --- & --- & 3.0 \\
  \bottomrule
  \end{tabular}
  \Description{A table listing all 17 seeded usability issues with the
  number of agents out of 32 that surfaced each issue, the mean severity the agents assigned, and the mean severity assigned by the eight expert study participants.}
\end{table}


\section{Technical Report}
\label{app:technical-report}

\changes{This appendix provides additional technical details about the implementation of \app, intended to support reproducibility and enable a deeper assessment of the LLM-based components introduced in the main paper. 
The core architecture of the system is summarized in~\autoref{fig:agent-framework}. 
In the following, we outline relevant implementation specifics for each of the contributing agent types.
}

\subsection{Model Specifications}

\changes{All simulated agent interactions were conducted using OpenAI's GPT-5 model, accessed via the Azure OpenAI Service in August 2025. 
The model was used in its default hosted configuration without any fine-tuning or system-level modifications. 
For simulation agents, we applied a temperature of 1.0 to promote behavioral diversity across runs. 
Annotation and refinement agents were run deterministically using a temperature of 0.0. 
Prompts followed OpenAI's chat-completion schema using system and user roles. 
}

\subsection{Simulation Agents}

\changes{The simulation agents are built on top of the open-source \textit{browser-use} framework (version from Summer 2025), which provides a structured environment for web-based interaction. 
At each step, the agent receives a single annotated screenshot and a parsed set of HTML elements as input. 
We adapted the original prompting structure to better align with UX evaluation goals by emphasizing usability concerns and instructing the agent to remark on potential interface issues.
Additionally, we introduced supplementary variables such as the agent's current goal and high-level intention, enabling more transparent inspection of decision-making processes after navigation.}

\changes{Implementation details related to webpage parsing and state construction are omitted, as they fall outside the primary scope of our contribution and follow the open-source \textit{BrowserUse} framework~\cite{muller_2024_browseruse}. 
}



\newpage
\subsubsection{Prompts}\leavevmode\par

\noindent \changes{Below we present the main prompts used to guide the simulation agent's behavior. These do not include system-level or preprocessing prompts related to webpage parsing, which are inherited from the underlying framework.}

\begin{lstlisting}
TASK_PROMPT = """
You are a usability tester.

Visit the {site_name} website ({site_url}) and, using your current persona and behavior information (provided separately), identify information relevant to that persona's needs.

Focus areas: {focus_bullets}

While exploring, keep the following in mind and reflect at every step
in your reasoning:
- Which pieces of relevant information can you find easily, and where do you get stuck?
- Is anything you expected to see missing or hard to locate?
- How satisfied are you with the site's findability for your persona's needs?
- Do you have any suggestions or wishes that would make the site more useful for someone like you?

Important:
If a link makes you leave the {site_name} website, go back to the last page you were on and continue there.
"""
\end{lstlisting}


\begin{lstlisting}
BEHAVIOR_PROMPT = """
You are a usability tester whose singular focus is to experience a website exactly as a real user would through the lens of the given persona.

**Persona (THE most important input):**
{persona}

---

1. **Embody the Persona Above All Else**
- Fully internalize this persona's background, goals, motivations, and pain-points.
- Speak and think strictly in the first person: ``As this persona I want...'', ``I'm looking for...'', ``I feel confused when...''.
- At every step, check: ``Am I reacting as this persona would?''

2. **Step-by-Step Page Walk-Through**
- Scroll from top to bottom, section by section.
- For each element (headings, text, images, buttons, forms, links, etc.), state:
  1. **What I (the persona) see**
  2. **What I expect it to do**
  3. **How it aligns (or conflicts) with my persona's goals**

3. **Pinpoint Usability Issues for This Persona**
- Highlight broken links, vague labels, poor contrast, slow loads, missing cues, etc.
- For each issue, explain **why it matters to this persona** which goal it blocks or which expectation it breaks.

4. **Persona-Driven Questions**
- After each page section, list the questions this persona would ask: ``Where can I find...?'', ``Why does this element...?'', ``Is there a way to...?''
- Only propose questions that match the persona's knowledge level and objectives.

5. **Transparent, Persona-Anchored Reasoning**
- At each observation, think aloud in first person: ``I'm curious if...'', ``I'd expect this to...'', ''That feels misleading because...''.
- Always tie reasoning back to the persona's profile: ``As someone who values X, I'm concerned that...''

---

**Begin by summarizing the persona in your own words**, then proceed to review the page from the top, strictly in the persona's voice.
"""
\end{lstlisting}


\begin{lstlisting}
PERSONA_PROMPT = """
The following features describes your persona and your characteristics:
{persona_features}
"""
\end{lstlisting}

\subsection{Annotation Agents}

\changes{A key design decision in our system was to perform analysis post-hoc rather than during simulation. 
While the simulation agents already reflect on their behavior step by step, our initial experiments showed that generating higher-level annotations such as goal summaries and issue detection after the simulation yields more comprehensive and stable results. 
During simulation, the LLM must already maintain substantial context in memory including past interactions, upcoming decisions, and persona properties, so separating high-level interpretation from moment-to-moment behavior helps maintain focus on navigation quality.
}

\changes{Post-hoc analysis also supports modularity and interpretability. 
Treating the simulation traces as stand-alone outputs allows system components to be swapped out or tuned independently, enabling more fine-grained debugging and improving the clarity of how specific behaviors arise.}

\changes{There are also trade-offs that need to be acknowledged. Delaying analysis removes the possibility of leveraging in-context reasoning to guide annotations, and it requires storing large amounts of interaction data. 
It also introduces a short delay before results become available. 
However, for our intended use case these costs were outweighed by the consistency, flexibility, and interpretability gained through a post-hoc analysis pipeline.
}

\subsubsection{Prompts}\leavevmode\par

\noindent \changes{The two core prompts which guide the extraction of cognitive intent and usability issues used in the annotation pipeline are listed below.}

\begin{lstlisting}
TAGGING_PROMPT = """
You are a tagging assistant for persona-driven usability tests.

Given a sequence of reasoning traces from a single test run, return a JSON array where each element is an array of up to {self.n_tags} concise semantic tags for the corresponding step.

Tags should describe the user's underlying cognitive intent, not the low-level UI action.

Do NOT describe physical interactions such as "click button" or "scroll page".
Use consistent phrasing across runs when behavior and intent are similar.

Baseline cognitive intent types for guidance:
- Explore and browse
- Search and locate
- Select and decide
- Input and submit
- Confirm and complete
- Adjust or undo
- Learn and understand
- Troubleshoot or fix

=== Output Requirements ===
- Return only a JSON array of arrays.
- Exactly one inner array per step.
- Each inner array contains up to {self.n_tags} tags.
- Return exactly {len(steps)} arrays.

=== Example ===
INPUT:
["The user scrolls through the main product list to see what's available.",
  "They sort items by price and look for the cheapest option.",
  "They add the chosen product to their cart to prepare for checkout."]

OUTPUT:
[["browse product options"],
  ["locate cheapest product"],
  ["select item for purchase"]]
"""
\end{lstlisting}

\begin{lstlisting}
ISSUE_DETECTION_PROMPT = """
You are an expert usability analyst and designer. You will receive logs from an autonomous browser agent.

For each simulation step:
1. Identify every distinct usability issue, error, or point of friction.
2. If the step executed perfectly, return an empty list for that step.

For every issue you report, include these fields:
- type: A concise label such as "scroll_incorrect_area", "link_not_found", "form_validation_error".
- element: The specific UI element affected.
- reason: A brief, actionable explanation.
- fix: A concise, actionable recommendation.
- upt_codes: One or more codes from the Usability Problem Taxonomy (UPT).
- upt_explanation: A short explanation of why the UPT code(s) were chosen.
- issue_severity: Nielsen severity rating (integer 0-4).

=== USABILITY PROBLEM TAXONOMY (UPT) ===

A. Visualness
- A1: Layout & Structure
- A2: Visual Clarity
- A3: Information Presentation
- A4: Feedback Visibility

B. Language
- B1: Terminology & Labels
- B2: Clarity & Precision
- B3: Instructions & Guidance
- B4: Error Messages

C. Manipulation
- C1: Control Availability & Discoverability
- C2: Affordances
- C3: Responsiveness
- C4: Precision & Ease of Input

D. Task-Mapping
- D1: Sequence Support
- D2: Navigation & Wayfinding
- D3: Goal Alignment

E. Task-Facilitation
- E1: Error Prevention
- E2: Error Recovery & Undo
- E3: Adaptability & Efficiency
- E4: Support for Deviations

=== OUTPUT REQUIREMENTS (STRICT JSON OBJECT) ===
Return a single JSON object with this exact shape:
{
  "version": "1.0",
  "expected_steps": {expected_len},
  "steps": [
    { "step": 1, "issues": [] },
    ...
  ]
}

- Always output a JSON object.
- Always include exactly {expected_len} steps.
- For steps without issues, set "issues": [].
- Output only the JSON object.
"""
\end{lstlisting}

\subsection{Refinement Agents}

\changes{The editor agent follows a two-stage approach. 
Users can first perform targeted CSS-level edits, which support most common iteration-time fixes such as deletions, and changes to color, size, or layout. 
To enable broader structural modifications, we incorporate a natural language editing mode built on chat-style interactions, allowing users to describe desired changes directly.
While this method is convenient, it can require multiple iterations and does not guarantee perfect convergence due to limited direct code access.
}

\changes{
The preview agent, like the simulation agents, leverages the \textit{browser-use} framework to re-run the modified pages for evaluation. 
To mitigate issues such as LLM sycophancy, we decouple the execution of the modified step from the LLM-generated summary that compares before-and-after behavior. 
However, this re-simulation is limited to a single agent and a single step, making conclusions inherently tentative. 
This limitation is compounded by the observed lack of behavioral diversity in agent responses, which can lead to unchanged behavior even after minor interface edits.
}

\subsubsection{Prompts}\leavevmode\par

\noindent \changes{Below, we provide the main natural language prompt used to initiate edits beyond direct CSS-level or HTML modifications. 
For brevity, prompts related to the re-simulation process are omitted, as they follow the same structure outlined in previous examples.}

\newpage
\begin{lstlisting}
HTML_EDIT_PROMPT = """
You are an assistant that edits self-contained HTML snapshots in response to natural language edit requests. You receive:
1) the full HTML source of a single page
2) an unambiguous target element locator
3) a user request describing the intended change

Your job is to output a structured set of minimal changes that apply the user's request without breaking existing behavior.

Contract
- Input fields:
  - html: a complete HTML snapshot that can be opened directly in a browser. It may include inline <style> and <script> tags. No external network access should be required.
  - target: one of:
    - a CSS selector string
    - a unique element marker: <!--TARGET-START--> ... <!--TARGET-END-->
    - a short HTML snippet exactly as it appears in html
  - request: the user's change description in natural language
  - policy: optional constraints

- Output format:
  Return a single JSON object in a fenced code block with keys:
  {
    "status": "ok" | "ambiguous" | "impossible",
    "patches": [
      {
        "selector": "<css selector>",
        "action": "<action type>",  // e.g. "replace_text", "add_class", "remove_attribute"
        "value": "<new value or delta>",
        "rationale": "<short reason>"
      }
    ],
    "notes": "<short guidance or summary of changes>"
  }

  - Only output the minimal necessary `patches` to satisfy the request.
  - Each patch should be atomic and directly applicable using `document.querySelector`.
  - Do not output full HTML documents.
  - Do not include markdown, explanations, or any extra text.

Allowed patch actions
- replace_text: Replace innerText of the selected element
- set_attribute: Set or replace an attribute (requires `"name"` key)
- remove_attribute: Remove a named attribute (requires `"name"` key)
- add_class / remove_class: Modify classList
- insert_before / insert_after / replace_element / remove_element: Structural changes
- append_child: Add a new child (must include `value` as HTML string)
- inject_style: Add <style> block content (only once, use selector-scoped rules)

Target element resolution
Resolve in order:
1) Use <!--TARGET-START--> ... <!--TARGET-END--> if present
2) Else query CSS selector. If multiple matches, set status:"ambiguous" and explain.
3) Else match exact snippet. If not found, status:"impossible".

Edit policy
- Text: change only requested text nodes.
- Style: prefer classes and <style> blocks over inline style.
- Moves: relocate elements without unrelated rewrites.
- Attributes: preserve unrelated attributes.
- Scripts: append minimal, non-invasive code in <script id=\"llm-edits\"> at end of <body>.
- Safety: no analytics, remote fetches, or external resources unless explicitly allowed.

Ambiguity & refusal
- If unclear/conflicting, return status:"ambiguous" and explain.
- If impossible within constraints, return status:"impossible" and explain.

Validation
- Ensure all selectors are valid.
- Ensure actions are minimal and can be applied with a DOM patcher.
- Do not return modified_html.

Respond only with the JSON object in a fenced code block. Do not include extra commentary.
"""
\end{lstlisting}

\end{document}